\documentclass[11pt]{article}

\usepackage[margin=1in]{geometry}
\usepackage{microtype}
\usepackage{graphicx}
\usepackage{amsmath}
\usepackage{iftex}

\newif\ifarxiv
\arxivtrue

\ifXeTeX
  \usepackage{fontspec}
  \usepackage{unicode-math}
  \defaultfontfeatures{Ligatures=TeX,Scale=MatchLowercase}

\else
  \usepackage{fontspec}
  \usepackage{unicode-math}
  \defaultfontfeatures{Ligatures=TeX,Scale=MatchLowercase}
  \IfFontExistsTF{Libertinus Serif}{%
    \setmainfont{Libertinus Serif}%
  }{%
    \setmainfont{TeX Gyre Termes}%
  }%
  \IfFontExistsTF{Libertinus Sans}{%
    \setsansfont{Libertinus Sans}[
      UprightFont=LibertinusSans-Regular.otf,
      BoldFont=LibertinusSans-Bold.otf,
      ItalicFont=LibertinusSans-Italic.otf,
      BoldItalicFont=LibertinusSans-Bold.otf,
      BoldItalicFeatures={FakeSlant=0.2}
    ]%
  }{%
    \setsansfont{TeX Gyre Heros}%
  }%
  \IfFontExistsTF{TeX Gyre Cursor}{%
    \setmonofont{TeX Gyre Cursor}[Scale=MatchLowercase]%
  }{%
    \setmonofont{Latin Modern Mono}[Scale=MatchLowercase]%
  }%
  \IfFontExistsTF{Libertinus Math}{%
    \setmathfont{Libertinus Math}[BoldFont=LibertinusMath-Regular.otf]%
  }{%
    \setmathfont{Latin Modern Math}%
  }%
\fi
\usepackage{textcomp}
\usepackage[dvipsnames,table]{xcolor}
\usepackage{xstring}
\usepackage{xurl}
\usepackage{ragged2e}
\usepackage{booktabs}
\usepackage{tabularx}
\usepackage{longtable}
\usepackage{array}
\usepackage{makecell}
\usepackage{placeins}
\usepackage{tikz}
\usetikzlibrary{patterns}
\usepackage{pgf-pie}
\usepackage{lscape}

\newif\ifeyeqrefcallouts
\eyeqrefcalloutsfalse
\ifdefined\HCode
  \eyeqrefcalloutsfalse
\fi
\ifeyeqrefcallouts
		  \usepackage{bigfoot}

		  \makeatletter
		  \renewcommand{\footnoterule}{%
		    \kern -3pt
		    \hrule width \linewidth height 0.4pt
		    \kern 4pt
		  }
		  
		  \newlength{\eyeqfnmarkwidth}
		  \newlength{\eyeqfngap}
		  \newlength{\eyeqfnindent}
		  \renewcommand{\@makefntext}[1]{%
		    \noindent
		    \hangindent=\eyeqfnindent
		    \hangafter=1
		    \makebox[\eyeqfnmarkwidth][r]{\@textsuperscript{\normalfont\@thefnmark}}%
		    \hspace{\eyeqfngap}%
		    \emergencystretch=1em\relax
		    #1%
		  }
		  \makeatother

	  \DeclareNewFootnote{B}
	  \MakeSorted{footnoteB}
\fi

\usepackage[hidelinks]{hyperref}
\ifdefined\HCode\else
  \usepackage{hyperxmp}
\fi

\let\eyeqorigpath\path
\renewcommand{\path}[1]{\begingroup\urlstyle{tt}\eyeqorigpath{#1}\endgroup}
\newcommand{\codeid}[1]{\path{#1}}

\definecolor{linkcolor}{HTML}{99479B}
\definecolor{citecolor}{rgb}{0,0.4,0}
\definecolor{urlcolor}{rgb}{0,0,0.65}
\definecolor{mycolor}{rgb}{0.122, 0.435, 0.698}
\definecolor{codebg}{HTML}{F8FAFC}
\definecolor{tighnari1}{RGB}{36,50,88}
\definecolor{tighnari2}{RGB}{40,114,70}
\definecolor{tighnari3}{RGB}{177,196,77}
\definecolor{tighnari4}{RGB}{231,199,54}
\definecolor{tighnari5}{RGB}{247,191,99}
\definecolor{tighnari6}{RGB}{238,176,175}
\definecolor{tighnari7}{RGB}{150,52,96}
\definecolor{tighnari8}{RGB}{99,31,102}
\definecolor{tighnari9}{RGB}{141,87,41}
\definecolor{tighnari10}{RGB}{1,137,157}

\hypersetup{
  pdfpagemode=UseNone,
  pdfstartview=FitH,
  unicode=true,
  pdflang={en-US},
  pdfdisplaydoctitle=true,
  pdftitle={Following Dragons: Code Review-Guided Fuzzing},
  pdfauthor={Viet Hoang Luu, Amirmohammad Pasdar, Wachiraphan Charoenwet, Toby Murray, Shaanan Cohney, Van-Thuan Pham},
  pdfkeywords={fuzzing, code review, developer intelligence, IJON, AFL++},
  colorlinks=true,
  linkcolor=linkcolor,
  urlcolor=urlcolor,
  citecolor=citecolor,
  bookmarksnumbered,
  breaklinks,
  bookmarksopen=false,
}

\usepackage[noabbrev,nameinlink]{cleveref}

\usepackage[inline]{enumitem} 

\definecolor{codebg}{HTML}{F8FAFC}
\newcommand{\eyeqMintedCacheDir}{_minted}
\IfFileExists{_minted/friendly.style.minted}{}{%
  \renewcommand{\eyeqMintedCacheDir}{../../_minted}%
}
\ifdefined\eyeqMintedFinalizeCache
  \usepackage[finalizecache,cachedir=\eyeqMintedCacheDir]{minted}
\else
  \usepackage[frozencache,cachedir=\eyeqMintedCacheDir]{minted}
\fi
\setminted{
  fontsize=\footnotesize,
  breaklines,
  breakanywhere,
  xleftmargin=0pt,
  xrightmargin=0pt,
  bgcolor=codebg,
  frame=lines,
  framesep=1mm,
  framerule=0.4pt,
  rulecolor=\color{black!25},
  breaksymbolleft=\raisebox{0.8ex}{\tiny\textcolor{black!35}{\ensuremath{\hookrightarrow}}}\,
}

\ifnum\shellescape=0\relax
  \renewcommand{\mintinline}[3][]{\texttt{\detokenize{#3}}}
\fi

\newcolumntype{Y}{>{\RaggedRight\arraybackslash}X}

\usepackage{needspace} 
\usepackage{etoolbox}
\pretocmd{\section}{\Needspace{5\baselineskip}}{}{}
\pretocmd{\subsection}{\Needspace{4\baselineskip}}{}{}
\pretocmd{\subsubsection}{\Needspace{3\baselineskip}}{}{}
\usepackage{tcolorbox}
\tcbuselibrary{skins,breakable}
\usepackage{caption}
\usepackage{subcaption}
\usepackage{colortbl}
\usepackage{mdframed}
\usepackage{pifont}
\usepackage[normalem]{ulem}

\usepackage{newfloat}
\makeatletter
\@ifundefined{listing}{%
  \DeclareFloatingEnvironment[
    fileext=lol,
    name=Listing,
    listname={List of Listings}
  ]{listing}%
}{}%
\makeatother

\definecolor{dkgray}{gray}{0.35}
\definecolor{vietgreen}{RGB}{0,100,0}

\newcommand{\codepath}[1]{\path{#1}}

\newcommand{\issuelink}[1]{\href{https://github.com/php/php-src/issues/#1}{#1}}

\providecommand{\textsubscript}[1]{\raisebox{-0.35ex}{\scriptsize #1}}

\DeclareRobustCommand{\EyeQNameItalic}{\textit{EyeQ}}
\DeclareRobustCommand{\EyeQName}{\EyeQNameItalic}
\providecommand{\eyeq}[1]{}
\providecommand{\eyeqhuman}[1]{}
\providecommand{\eyeqllm}[1]{}
\renewcommand{\eyeq}[1]{\EyeQName}
\renewcommand{\eyeqhuman}[1]{\EyeQName\textsubscript{\textup{Human}}}
\renewcommand{\eyeqllm}[1]{\EyeQName\textsubscript{\textup{LLM}}}

\newcommand{\mylno}[1]{%
  \makebox[1.4em][r]{\tiny\color{dkgray}#1}\hspace{0.6em}%
}
\makeatletter
\newcommand{\codeanchor}[1]{\Hy@raisedlink{\hypertarget{#1}{}}}
\makeatother

\definecolor{softred}{HTML}{6E5A5A}
\newcommand{\reduline}[1]{\textcolor{softred}{\uline{#1}}}
\DeclareRobustCommand{\capcode}[1]{\texttt{#1}}
\newcommand{\reply}{%
  \par\noindent
  \raisebox{0.25em}{\textcolor{black!50}{\rule{0.4pt}{1.2em}}}%
  \hspace{-0.4pt}%
  \raisebox{0.9em}{\textcolor{black!50}{\rule{0.8em}{0.4pt}}}%
  \hspace{0.6em}%
}

\definecolor{contribbg}{HTML}{FFF7ED} 
\definecolor{contribfg}{HTML}{7C2D12} 
\definecolor{insightbg}{HTML}{F2F7FF} 
\definecolor{insightfg}{HTML}{2B4C7E} 
\definecolor{flowbg}{HTML}{F0FDF4}    
\definecolor{flowfg}{HTML}{166534}    

\newcommand{\callouttitle}[1]{\textbf{\textcolor{insightfg}{#1}}}
\newcommand{\boxtitle}[2]{\textbf{\textcolor{#1}{#2}}}
\newcommand{\clabel}[1]{\textcolor{contribfg}{\textbf{C#1}}}

\newtcolorbox{contribbox}{
  enhanced, breakable,
  colback=contribbg, colframe=contribfg,
  boxrule=0pt, frame hidden,
  borderline west={2pt}{0pt}{contribfg},
  left=8pt,right=8pt,top=6pt,bottom=6pt,
}
\newtcolorbox{insightbox}{
  enhanced, breakable,
  colback=insightbg, colframe=insightfg,
  boxrule=0pt, frame hidden,
  borderline west={2pt}{0pt}{insightfg},
  left=8pt,right=8pt,top=6pt,bottom=6pt,
}
\newtcolorbox{workflowbox}{
  enhanced, breakable,
  colback=flowbg, colframe=flowfg,
  boxrule=0.4pt, arc=2pt,
  left=8pt,right=8pt,top=6pt,bottom=6pt,
}
\newtcolorbox{reviewquotebox}{
  enhanced, breakable,
  width=\linewidth,
  colback=white,
  colframe=black!30,
  boxrule=0.5pt,
  arc=2pt,
  left=8pt,right=8pt,top=6pt,bottom=6pt,
  fontupper=\footnotesize\RaggedRight,
}
\newtcolorbox{reviewquoteboxfloat}{
  enhanced,
  width=\linewidth,
  colback=white,
  colframe=black!30,
  boxrule=0.5pt,
  arc=2pt,
  left=8pt,right=8pt,top=6pt,bottom=6pt,
  fontupper=\footnotesize\RaggedRight,
}

\newcommand{\eyeqBibBackend}{bibtex}
\usepackage[
  backend=\eyeqBibBackend,
  bibstyle=numeric,
  citestyle=numeric-comp,
  citetracker=true,
  sorting=none,
  sortcites=true,
  mincitenames=1,
  maxcitenames=1,
  maxbibnames=99,
  url=true,
  isbn=false,
  backref=true,
  doi=false,
  eprint=false,
  hyperref=true,
  bibencoding=utf8
]{biblatex}
\providecommand{\Textcite}[1]{\textcite{#1}}
\providecommand{\Creflabelformat}[2]{\creflabelformat{#1}{#2}}
\crefname{appendix}{\textbf{Appendix}}{\textbf{Appendices}}
\Crefname{appendix}{\textbf{Appendix}}{\textbf{Appendices}}
\creflabelformat{appendix}{#2\textbf{#1}#3}
\Creflabelformat{appendix}{#2\textbf{#1}#3}
\crefname{section}{\textbf{Section}}{\textbf{Sections}}
\Crefname{section}{\textbf{Section}}{\textbf{Sections}}
\creflabelformat{section}{#2\textbf{#1}#3}
\Creflabelformat{section}{#2\textbf{#1}#3}
\crefname{subsection}{\textbf{Section}}{\textbf{Sections}}
\Crefname{subsection}{\textbf{Section}}{\textbf{Sections}}
\creflabelformat{subsection}{#2\textbf{#1}#3}
\Creflabelformat{subsection}{#2\textbf{#1}#3}
\crefname{subsubsection}{\textbf{Section}}{\textbf{Sections}}
\Crefname{subsubsection}{\textbf{Section}}{\textbf{Sections}}
\creflabelformat{subsubsection}{#2\textbf{#1}#3}
\Creflabelformat{subsubsection}{#2\textbf{#1}#3}
\crefname{equation}{\textbf{Equation}}{\textbf{Equations}}
\Crefname{equation}{\textbf{Equation}}{\textbf{Equations}}
\creflabelformat{equation}{#2\textbf{(#1)}#3}
\Creflabelformat{equation}{#2\textbf{(#1)}#3}
\crefname{figure}{\textbf{Figure}}{\textbf{Figures}}
\Crefname{figure}{\textbf{Figure}}{\textbf{Figures}}
\creflabelformat{figure}{#2\textbf{#1}#3}
\Creflabelformat{figure}{#2\textbf{#1}#3}
\crefname{table}{\textbf{Table}}{\textbf{Tables}}
\Crefname{table}{\textbf{Table}}{\textbf{Tables}}
\creflabelformat{table}{#2\textbf{#1}#3}
\Creflabelformat{table}{#2\textbf{#1}#3}
\crefname{listing}{\textbf{Listing}}{\textbf{Listings}}
\Crefname{listing}{\textbf{Listing}}{\textbf{Listings}}
\creflabelformat{listing}{#2\textbf{#1}#3}
\Creflabelformat{listing}{#2\textbf{#1}#3}
\crefname{lstlisting}{\textbf{Listing}}{\textbf{Listings}}
\Crefname{lstlisting}{\textbf{Listing}}{\textbf{Listings}}
\creflabelformat{lstlisting}{#2\textbf{#1}#3}
\Creflabelformat{lstlisting}{#2\textbf{#1}#3}

\definecolor{calloutbg}{HTML}{F6FAFF}   
\definecolor{calloutbd}{HTML}{2B4C7E}   
\definecolor{callouttl}{HTML}{E6F0FF}   

\tcbset{
eyeqcallout/.style={
      enhanced,
      breakable,
      colback=calloutbg,
      colframe=calloutbd,
      boxrule=0.5pt,
      arc=2pt,
      left=7pt,right=7pt,top=6pt,bottom=6pt,
      borderline west={2.2pt}{0pt}{calloutbd}, 
      colbacktitle=callouttl,
      coltitle=calloutbd,
      fonttitle=\bfseries\small,
      attach boxed title to top left={xshift=6pt,yshift=-1.2mm},
      boxed title style={boxrule=0pt,arc=2pt,left=4pt,right=4pt,top=2pt,bottom=2pt},
    }
}

\newtcolorbox{findingbox}[1]{eyeqcallout,title={#1}}

\newtcolorbox{promptbox}[1][]{
  enhanced,
  width=\linewidth,
  colback=white,
  colframe=black!40,
  boxrule=0.6pt,
  arc=2.5pt,
  left=6pt,right=6pt,top=4pt,bottom=4pt,
  before upper={\scriptsize\RaggedRight\sloppy\setlength{\parindent}{0pt}},
  after upper={\fussy\normalsize},
  fonttitle=\bfseries\footnotesize,
  coltitle=black,
  #1
}

\newcommand{\listingcaptionbelow}[2]{%
  \par\vspace{-0.25em}%
  {\captionsetup{justification=centering,singlelinecheck=true,hypcap=false}%
    \captionof{listing}{#1}%
    \label{#2}}%
  \par\vspace{0.75em}%
}

\title{Following Dragons: Code Review-Guided Fuzzing}
\author{%
\begin{tabularx}{\linewidth}{@{}>{\centering\arraybackslash}X>{\centering\arraybackslash}X@{}}
\begin{tabular}{@{}c@{}}
Viet Hoang Luu\\[0.15em]
{\small\href{mailto:viethoangl@student.unimelb.edu.au}{\texttt{viethoangl@student.unimelb.edu.au}}}
\end{tabular}
&
\begin{tabular}{@{}c@{}}
Amirmohammad Pasdar\\[0.15em]
{\small\href{mailto:Amirmohammad.Pasdar@unimelb.edu.au}{\texttt{Amirmohammad.Pasdar@unimelb.edu.au}}}
\end{tabular}
\\[1.2em]
\begin{tabular}{@{}c@{}}
Wachiraphan Charoenwet\\[0.15em]
{\small\href{mailto:wcharoenwet@student.unimelb.edu.au}{\texttt{wcharoenwet@student.unimelb.edu.au}}}
\end{tabular}
&
\begin{tabular}{@{}c@{}}
Toby Murray\\[0.15em]
{\small\href{mailto:toby.murray@unimelb.edu.au}{\texttt{toby.murray@unimelb.edu.au}}}
\end{tabular}
\\[1.2em]
\begin{tabular}{@{}c@{}}
Shaanan Cohney\\[0.15em]
{\small\href{mailto:shaanan@cohney.info}{\texttt{shaanan@cohney.info}}}
\end{tabular}
&
\begin{tabular}{@{}c@{}}
Van-Thuan Pham\\[0.15em]
{\small\href{mailto:thuan.pham@unimelb.edu.au}{\texttt{thuan.pham@unimelb.edu.au}}}
\end{tabular}
\\[1.5em]
\multicolumn{2}{@{}c@{}}{\textit{The University of Melbourne}}%
\end{tabularx}%
}
\date{}

\newcommand{\dragon}{}
\ifLuaTeX
  \IfFontExistsTF{Noto Emoji}{%
    \newfontfamily\emojifont{Noto Emoji}[
      Renderer=Harfbuzz,
      NoSubset,
      BoldFont={Noto Emoji},
      ItalicFont={Noto Emoji},
      BoldItalicFont={Noto Emoji}
    ]%
    \renewcommand{\dragon}{{\emojifont \char"1F409}}%
  }{%
    \renewcommand{\dragon}{}%
  }
\else
  \IfFileExists{fontawesome5.sty}{%
    \usepackage{fontawesome5}%
    \renewcommand{\dragon}{\faDragon}%
  }{}
\fi

\begin{document}
\maketitle

\begin{abstract}
Modern fuzzers scale to large, real-world software but often fail to exercise the program states developers consider most fragile or security-critical. Such states are typically deep in the execution space, gated by preconditions, or overshadowed by lower-value paths that consume limited fuzzing budgets. Meanwhile, developers routinely surface risk-relevant insights during code review, yet this information is largely ignored by automated testing tools.

We present EyeQ, a system that leverages developer intelligence from code reviews to guide fuzzing. EyeQ extracts security-relevant signals from review discussions, localizes the implicated program regions, and translates these insights into annotation-based guidance for fuzzing. The approach operates atop existing annotation-aware fuzzing, requiring no changes to program semantics or developer workflows.

We first validate EyeQ through a human-guided feasibility study on a security-focused dataset of PHP code reviews, establishing a strong baseline for review-guided fuzzing. We then automate the workflow using a large language model with carefully designed prompts. EyeQ significantly improves vulnerability discovery over standard fuzzing configurations, uncovering more than 40 previously unknown bugs in the security-critical PHP codebase.

\end{abstract}

\section{Introduction}
{\begin{center}
\textit{Code review is often where developers mark ``here be dragons.'' Conventional fuzzing rarely reads the map. }\end{center}}

\vspace{0.05em}
\noindent\rule{\linewidth}{0.3pt}
\vspace{0.1em}

This paper shows how to extract developers' risk-relevant insights from code review---about fragile, complex, or security-critical logic---and use them to steer fuzzers toward high-risk program states.

Fuzzing occupies a distinctive point in the space of vulnerability discovery: it scales to real-world software, operates directly on executable artifacts, and continues to uncover vulnerabilities even in heavily scrutinized systems \cite{manes2019art}. 

However, fuzzers often fail to test the program states developers worry about most---because they are \begin{enumerate*}[label=(\arabic*)] \item often deep in the state graph; \item may be gated by \textit{blockers} (preconditions that must be satisfied); \item and fuzzers spend valuable time on less valuable parts of the space that they could instead spend targeting higher-value states \end{enumerate*}. As a result, vulnerabilities continue to evade state-of-the-art fuzzers.

This returns us to the value of incorporating additional signals to guide fuzzers. If a state is reachable \textit{ab initio}, a fuzzer has some probability of reaching the risky state on its own. However, in practice, time is a major constraint on vulnerability discovery: \emph{where} a fuzzer spends its time matters. Therefore we postulate that prioritizing states that developers themselves flag as risky could yield substantial real-world gains, improving the likelihood that limited fuzzing resources will uncover high-impact bugs.

\begin{insightbox}
\callouttitle{Key postulate.}
\textbf{Developers leave clues about where code is brittle; fuzzers can use those clues}. This does not magically bypass every blocker, but it could drive a fuzzer to spend more of its limited time in the \textbf{parts of the space where failures are most likely---and most costly}.
\end{insightbox}

A natural question, then, is which forms of developer-produced knowledge are best suited to this role. Software artifacts such as requirements documents, specifications, user manuals, and inline comments can all encode human insight, but in practice they are often incomplete or unavailable, and quickly become stale as code evolves. 

Code reviews differ from many other development artifacts in two important respects. First, they are tightly coupled to code changes and diffs---the moments when new logic is introduced and existing invariants are most likely to be disrupted. 
Second, code review often is explicit in focus on correctness, edge cases, and bugs.

In this work, we therefore focus on the exemplar of \textit{code reviews} as our source of human intelligence to guide fuzzing.

Leveraging code-reviews for fuzzing requires us to overcome a series of practical challenges, for each of which we propose a solution and evaluate its efficacy.  This leads to our core challenge, along with our proposed workflow:

\begin{workflowbox}
\callouttitle{Workflow.}
The greatest overarching challenge is in transforming unstructured review comments (written in natural language) into guidance \textit{that can be consumed by a fuzzing engine}.
We address this through our workflow that:
\begin{enumerate*}
  \item \textit{identifies security-relevant code reviews},
  \item \textit{localizes} the program components referenced by those reviews, and
  \item converts review insights into \textbf{\textit{program annotations}} that introduce additional guidance signals for fuzzing
\end{enumerate*}.
Our workflow is designed to operate atop fuzzing infrastructures that support annotation-based guidance, such as IJON \cite{aschermann2020ijon}, which enables the exploration of deep program state spaces through developer-provided annotations.
\end{workflowbox}

We first validate our proposed approach through a proof-of-concept study with a human analyst adding annotations. To do so we leverage an existing security-focused dataset of PHP code reviews \cite{charoenwet2024empirical} wherein developers raise many security concerns that are not addressed in subsequent commits. This produces a potential target-rich environment for our validation experiments.

For each review comment, we hand-analyze the associated pull request to localize the code regions implicated by the developer's concern. We then annotate the code with semantically appropriate \textbf{\textit{program annotations}}, and fuzz using AFL++ with IJON-guidance enabled.

The results from this were promising enough---41 previously unknown bugs in PHP---to validate the initial hypothesis. These results also provide an accompanying baseline against which to evaluate the automated approach. We then proceed to the automated approach (in which we invoke an LLM for manual tasks). The LLM here is not fundamental to capitalizing on our insights---rather it helps minimize human effort and to make the trade-offs worthwhile.

Finally, we evaluate the automated approach and show that our system significantly improves vulnerability discovery compared to baseline fuzzing configurations and provides for new directions in augmenting automated testing with human intelligence.\\

\begin{contribbox}
\boxtitle{contribfg}{Highlighted Contributions.}
\begin{description}[style=nextline,leftmargin=2.6em,labelwidth=2.2em,itemsep=2pt,topsep=2pt]
  \item[\clabel{1}] We identify code reviews as a useful and previously underexploited source of human intelligence for guiding fuzzing.
  \item[\clabel{2}] We propose an abstract approach for transforming code review insights into annotation-based fuzzing guidance.
  \item[\clabel{3}] We operationalize our approach with both a human-guided feasibility study and subsequently a fully automated LLM-based system to realize this workflow at scale---evaluating both elements of our approach and the combined whole.
  \item[\clabel{4}] We discover 46 previously unknown bugs in the large, security-critical PHP project. 
\end{description}
\end{contribbox}

\ifLuaTeX\vspace*{1.0em}\fi
\section{Background and Motivating Example}
We begin with some necessary preliminaries for readers unfamiliar with fuzzing, where fuzzing gets stuck, and further, with typical development practices as of writing.

\subsection{Fuzzing and Fuzz Blockers}

Fuzzing is an automated testing technique that generates inputs and executes them against a target program to expand coverage and uncover abnormal behaviors such as crashes and memory safety violations. \emph{Code coverage–guided greybox fuzzing} (CGF)--as implemented in widely used open-source tools such as AFL/AFL++ \cite{fioraldi2020afl++}, LibAFL \cite{fioraldi2022libafl}, libFuzzer, and Honggfuzz--is among the most successful approaches in practice. Unlike blackbox fuzzing, which treats the program as an opaque entity, or whitebox techniques such as symbolic execution, which attempt exhaustive reasoning about program semantics, greybox fuzzing strikes a pragmatic balance.

A central idea in CGF is \emph{feedback-guided fuzzing}, in which the fuzzer adapts its input generation strategy based on observed execution behavior. In practice, code coverage serves as the primary feedback signal: transitions between basic blocks are tracked to approximate control-flow diversity while mitigating path explosion. In representative CGF tools such as AFL, the target program is instrumented to update a shared coverage bitmap, mapping executed edges to pseudo-random indices with discretized hitcounts. This design enables fast bitmap comparisons via bitwise operations, sustaining thousands of executions per second.

CGF has demonstrated strong empirical effectiveness, including sustained success in large-scale deployments such as OSS-Fuzz \cite{ossfuzz_blogpost}, and it scales well to large and complex codebases where more precise analyses suffer from state explosion and prohibitive computational cost \cite{baldoni2018survey}.

Despite its success, \Textcite{gao2023beyond} identified several recurring \textit{blockers} that prevent fuzzers from reaching deeper program states. These obstacles arise along both \emph{untainted} and \emph{tainted} paths. 

A \emph{tainted path} is one in which key branches are controlled by bytes derived from the fuzzer’s input--i.e., input-tainted data flows into relevant conditionals--so progress requires generating inputs that satisfy semantic checks, including magic values, structured parsing constraints, or checksums. In contrast, \emph{untainted paths} are primarily gated by factors not directly influenced by input bytes, such as missing or incomplete drivers or harnesses, configuration and environment state, or scheduling behavior.

For untainted paths, progress has largely come from automating fuzz driver generation and related harnessing techniques, which remain an active area of research \cite{babic2019fudge, ispoglou2020fuzzgen, chen2023hopper}. For tainted paths, most approaches instead focus on strengthening the fuzzer itself, for example through improved mutation strategies \cite{pham2019smart, aschermann2019nautilus, aschermann2019redqueen}, richer feedback signals \cite{gan2020greyone, fioraldi2021use}, or more effective corpus management and scheduling \cite{she2022effective, bohme2020boosting}. Orthogonal to both directions, human-in-the-loop fuzzing \cite{shoshitaishvili2017rise, fang2024ddgf, gao2025insightql, aschermann2020ijon} demonstrates that developer insight can be highly effective in steering input generation and exploration toward otherwise hard-to-reach program behaviors.

However, existing human-in-the-loop approaches typically require significant manual effort, such as writing grammars, providing interactive feedback, or designing specialized guidance mechanisms. While effective, these methods do not scale well and introduce practical barriers to adoption.

\begin{insightbox}
\callouttitle{Key question.} Can we leverage human intelligence to guide fuzzing without requiring additional human effort during the fuzzing process?
\end{insightbox}

\subsection{Annotation-Aware Fuzzing}

Annotation-aware fuzzing, introduced by \Textcite{aschermann2020ijon}, augments coverage-guided fuzzing with semantic feedback to overcome a key limitation of edge coverage: its inability to reflect progress toward deep, vulnerability-prone program states. Many real-world bugs depend on subtle invariants or stateful logic that no longer produce new coverage once surrounding code is explored.

{
IJON addresses this by exposing user-defined program signals via lightweight instrumentation macros. Annotations such as \mintinline{c}{IJON_SET(x)} report numeric program state, while \mintinline{c}{IJON_MAX(x)} and \mintinline[breaklines,breakanywhere]{c}{IJON_MIN(x)} reward monotonic progress; discrete execution phases can be captured using \mintinline{c}{IJON_STATE()}. These signals are treated as first-class feedback, allowing the fuzzer to prioritize semantically meaningful executions even when coverage does not increase.
}

However, IJON assumes manual insertion of annotations by experts, which limits scalability. Our work builds on the same principle but removes this bottleneck by automatically extracting semantic guidance from code reviews.

\subsection{Modern Code Review}

Modern Code Review (MCR) has evolved into a lightweight, tool-supported, and socially embedded practice, largely replacing the formal, synchronous inspection models of early software engineering \cite{bacchelli2013expectations,fagan2011design,rigby2013convergent}. In contrast to meeting-driven reviews, MCR emphasizes asynchronous discussion, rapid iteration, and tight integration with version control systems.

Beyond coordinating collaboration, MCR serves as a rich forum for articulating developer reasoning about correctness, design trade-offs, and security risks. Reviewers routinely flag logic errors, unsafe assumptions, fragile invariants, and error-handling issues through inline comments and threaded discussions \cite{bacchelli2013expectations,mantyla2008types,balachandran2013reducing}. Empirical studies show that security-relevant reasoning is often expressed implicitly--as discussions of coding weaknesses or questionable logic rather than explicit vulnerability reports \cite{charoenwet2024empirical}. For instance, analyses of OpenSSL and PHP reviews find that weakness-related discussions occur 21–33 times more frequently than direct mentions of exploitable flaws, yet collectively span 35 of 40 CWE-699 categories \cite{charoenwet2024empirical}.

\begin{insightbox}
\callouttitle{Key insight.} Despite its effectiveness at eliciting high-quality human insight, MCR provides no systematic mechanism for reusing this knowledge beyond the review process itself. Many identified concerns remain partially mitigated, deferred, or unresolved, and patches with outstanding security issues are sometimes merged \cite{charoenwet2024empirical}. Consequently, valuable understanding of risky program states, fragile assumptions, and subtle behaviors is rarely propagated into later testing and validation phases. This motivates our approach: rather than treating code reviews solely as communication artifacts, we view them as a scalable source of semantic guidance that can be automatically extracted and translated into actionable feedback for dynamic analysis approaches like fuzzing.
\end{insightbox}

\vspace{0.8em}

\subsection{Motivating Example: A Missed Stack Overflow in PHP Fibers}

To illustrate the limitations of existing CGF fuzzers and the value of reviewer-guided semantic feedback, we examine a previously unknown bug in the Fibers component of the PHP runtime that was not discovered by continuous fuzzing prior to our work. PHP is a widely deployed scripting language that powers millions of websites worldwide. Fibers, introduced in PHP 8.1, provide lightweight user-space concurrency by allowing execution contexts to be paused and resumed with independent stacks.

The vulnerability (GitHub issue \#20483~\footnote{\url{https://github.com/php/php-src/issues/20483}}) manifests as a stack overflow in zend\_fiber\_execute() when the fiber.stack\_size configuration option is set to certain atypical values (e.g., "9690x-D", as shown in our proof-of-concept and \Cref{lst:fiber_overflow}). In particular, the bug is triggered when the configured stack size is malformed or unusually small yet still passes validation. Under these conditions, fiber execution can lead to unchecked stack growth during function calls or context switching, ultimately resulting in a stack overflow detected by AddressSanitizer and a process crash. The issue was subsequently fixed by strengthening the validation and handling of the fiber stack size to prevent unsafe configurations.

{
\begin{minted}[
  breaklines,
  breakanywhere,
  escapeinside=@@,
  xleftmargin=0.0em,
  xrightmargin=0.6em
]{php}
@\mylno{1}@<?php
@\mylno{2}@class A {
@\mylno{3}@    function __destruct() {
@\mylno{4}@        ini_set("fiber.stack_size", "9690x-D");
@\mylno{5}@        $fiber = new Fiber(function() {});
@\mylno{6}@        try {
@\mylno{7}@            $fiber->start();
@\mylno{8}@        } catch (Throwable $e) {}
@\mylno{9}@    }
@\mylno{10}@}
@\mylno{11}@new A;
@\mylno{12}@/* AddressSanitizer: stack-overflow ... */
\end{minted}
\listingcaptionbelow{PHP code that triggers a Stack Overflow by configuring a Malformed Fiber Stack size.}{lst:fiber_overflow}
}

Despite PHP having been integrated into OSS-Fuzz for several years and continuously fuzzed with a diverse set of manually curated test cases (as shown in Listing \ref{lst:fiber_close_enough_test}) and automatically generated seeds, this bug remained undiscovered until our analysis. This highlights a fundamental limitation of existing fuzzers: the vulnerability is not exposed by covering new control-flow paths, but by generating specific numeric values for a configuration parameter that induce unsafe runtime behavior. Coverage-guided fuzzers are generally ineffective in such scenarios, as mutating existing seeds rarely produces the precise values required to trigger the fault.

{
\begin{minted}[
  breaklines,
  breakanywhere,
  escapeinside=@@,
  xleftmargin=0.0em,
  xrightmargin=0.6em
]{php}
@\mylno{1}@<?php
@\mylno{2}@/* Zend/tests/fibers/negative_stack_size.phpt */
@\mylno{3}@ini_set("fiber.stack_size", "-1");
@\mylno{4}@/* expected: invalid stack size warning */
@\mylno{5}@/* Zend/tests/fibers/gh10249.phpt */
@\mylno{6}@ini_set("fiber.stack_size", "");
@\mylno{7}@/* expected: Fiber stack size is too small */
\end{minted}
\listingcaptionbelow{Existing Fiber Stack-Size Tests Exercising the Same Configuration Path}{lst:fiber_close_enough_test}
}

In principle, incorporating data-flow–aware feedback could help preserve test cases that exercise different stack size values. In practice, however, such approaches are often ineffective or impractical without prior knowledge of where value diversity is semantically meaningful. In this setting, effective tracking would need to target specific Fiber initialization and execution routines that propagate the stack size parameter. 

\begin{figure}[t]
\centering
\begin{reviewquoteboxfloat}
\footnotesize
\textbf{twose} (Jun 3, 2021):

\vspace{0.4em}

\reply If a large chunk of memory were allocated on the stack it could exceed 
the size of a single page.

I think what we need to guard against is 
\reduline{recursive calls rather than big memory on the stack}, 
and I did not see the code for multi-page memory protection in other coroutine 
projects. Maybe you can specify the reference of this idea, it is new to me. 
The cost of protecting multiple pages is far greater than the benefit, and we 
will never know how much is appropriate. Perhaps 
\reduline{the most common case is to alloc a large buffer on the stack, although 
it may push the pointer far beyond the stack boundary}; 
we always write from the beginning, don’t we? 

About symmetric and asymmetric, maybe you are entangled in these 
concepts, I only focused on its actual behaviour, I think these concepts are 
defined based on the behaviour of specified technological. I don't quite 
understand what you are worried about. And Swoole has never stated what kind 
of model its coroutine is, it do not care.
\end{reviewquoteboxfloat}

\caption{Code review discussion on fiber stack protection strategies in PHP.
Underlines highlight developer reasoning and domain knowledge expressed during the review.}
\label{fig:stack-overflow-fiber}
\end{figure}

Our approach addresses these challenges by systematically extracting semantic guidance from code review discussions. In this example, review comments (shown in \Cref{fig:stack-overflow-fiber}) implicitly highlighted concerns about Fiber stack sizing and safety assumptions, enabling identification of the relevant program locations. We leverage this information to inject targeted annotations that track and preserve meaningful changes to the stack size parameter. As a result, an annotation-aware fuzzer like IJON can generate inputs that trigger the stack overflow, leading to the discovery of this vulnerability and many others. The following sections explain how these techniques work.

\section{Code Review–Guided Fuzzing}
\label{sec:workflow}

\providecommand{\codeid}[1]{\texttt{#1}}

\begin{figure*}[t]
  \centering
  \includegraphics[width=\linewidth]{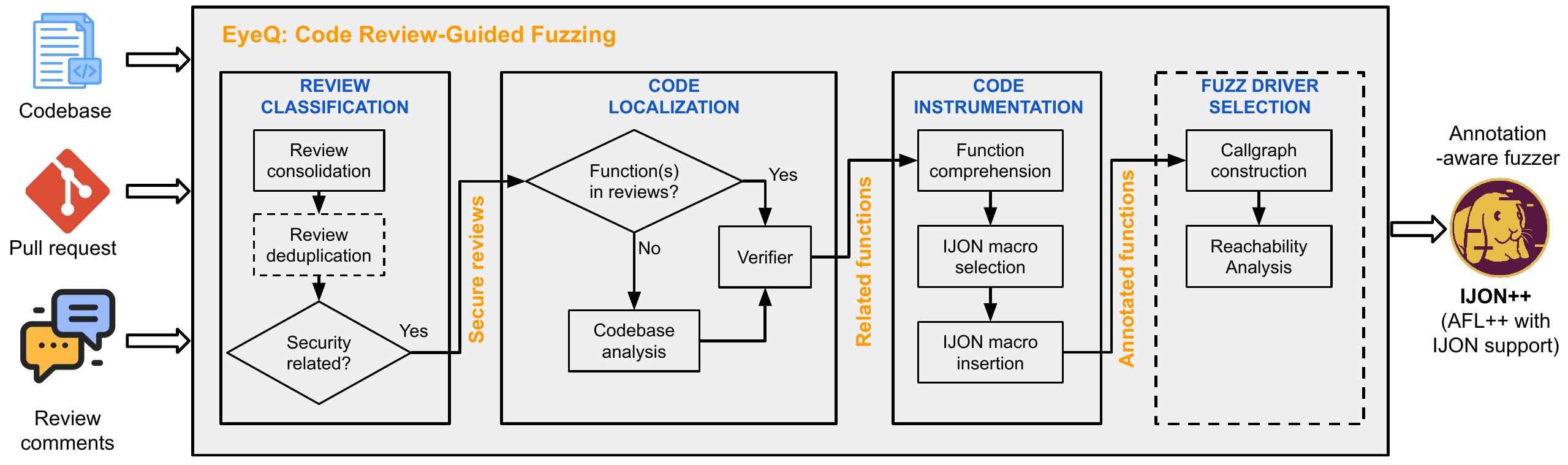}
  \caption{\textbf{End-to-end workflow of code review-guided fuzzing}. The pipeline transforms security-relevant reviews into localized code annotations and uses annotation-aware fuzzing to guide exploration toward vulnerability-prone program behaviors.} \label{fig:eyeq}
\end{figure*}


\Cref{fig:eyeq} presents the complete workflow of \eyeq{}, our novel code-review–guided fuzzing approach. The design is modular and abstract, allowing it to be instantiated either by human analysts (denoted as \eyeqhuman{}) or by a fully automated reasoning system based on large language models (denoted as \eyeqllm{}), while sharing the same underlying principles. The workflow operates within a standard development environment in which source code, pull requests, and review discussions are available, regardless of whether the repository is public or private.

Given a pull request that introduces a new feature or a bug fix and its associated review comments, \eyeq{} consists of four stages that transform review feedback into annotated code suitable for fuzzing.

In the first stage (Review Classification), a human analyst or an automated tool examines review comments to determine whether any are security related. If such comments are identified, the workflow proceeds to the second stage (Code Localization), which maps the selected comments to the corresponding regions of the codebase. The third stage (Code Instrumentation) then selects suitable variables or expressions, determines appropriate annotations (e.g., IJON annotations), and inserts them into the code.

To ensure that the modified code is reachable--particularly in projects with multiple fuzz drivers--the workflow may optionally include fuzz driver reachability analysis. This analysis can be performed manually through debugging and tracing or automatically using call-graph–based techniques.

Finally, the annotated codebase is compiled and fuzzed using an annotation-aware fuzzer, such as IJON~\cite{aschermann2020ijon}, with an appropriate seed corpus under a standard fuzzing setup. The additional semantic signals introduced by the annotations guide the fuzzer toward vulnerability-prone regions, increasing the likelihood of uncovering hard-to-find bugs and vulnerabilities in areas that are of particular interest to developers.

In the following subsections, we describe these stages in greater detail and demonstrate how they are applied to uncover the vulnerability in the motivating example.

\subsection{Stage 1: Review Classification}

For each pull request under analysis, review comments may appear in multiple locations, including general discussion threads, inline review comments attached to specific code changes, and formal review submissions. These artifacts capture complementary forms of developer and reviewer reasoning, ranging from high-level design concerns to line-level correctness checks and explicit acceptance or rejection decisions. In the case of a human analyst (\eyeqhuman{}), all such comments must be read and interpreted manually. In contrast, for an automated workflow such as \eyeqllm{}, these comments are first consolidated--and deduplicated when necessary--into a unified review corpus, while preserving relevant metadata such as pull request identifiers, file paths, and diff context when available.

\subsubsection{Principles of classifying reviews} To determine whether a review corpus contains security-related discussion and should therefore be marked as security relevant, we follow the methodology proposed by \Textcite{charoenwet2024empirical}. As observed by \Textcite{charoenwet2024empirical}, reviewers may not always have sufficient security expertise to identify concrete vulnerabilities or articulate explicit attack scenarios. Instead, they often discuss coding weaknesses that could potentially lead to security issues. For example, reviewers may raise concerns about input validation, which can result in vulnerabilities such as buffer overflows. Accordingly, a review is considered security related if reviewers explicitly mention security vulnerabilities or their security consequences (e.g., when reviewers have security expertise), or if they express concern about any of the 40 coding weaknesses listed in the widely used CWE-699 taxonomy\footnote{\url{https://cwe.mitre.org/data/definitions/699.html}}, which may indicate potential security issues.

\subsubsection{Example} Following these principles to the motivating example, we identify the review shown in \Cref{fig:stack-overflow-fiber} as security relevant. In this review comment, reviewers debate how fibers should guard against stack exhaustion, contrasting recursive call patterns with large on-stack allocations that could push the stack pointer far beyond the stack boundary. Although framed as an implementation-level concern and without explicit mention of potential vulnerabilities, the discussion implicitly describes a memory-safety risk consistent with \href{https://cwe.mitre.org/data/definitions/1218.html}{Memory Buffer Errors (CWE-1218)}
 and, more specifically, an \href{https://cwe.mitre.org/data/definitions/787.html}{Out-of-Bounds Write (CWE-787)}.

\subsubsection{Automation with LLM} In \eyeqllm{}, we automate the classification process by following the aforementioned principles and adopting a two-stage classification approach with suitable prompt design. The first stage, using the prompt shown in \Cref{fig:stage1-call1-prompt}, performs coarse filtering and category selection: given a review comment and a compact list of CWE-699 categories, the model determines whether the comment is security relevant and, if so, identifies a small set of candidate categories. This stage prioritizes recall and allows an explicit \texttt{uncertain} outcome to ensure that borderline cases are retained for further analysis. 

	\begin{figure}[t]
	  \centering
		  \begin{promptbox}
		  \textbf{System role.} You are a senior application security engineer reviewing developer comments in pull requests.

  \vspace{0.25em}
  \textbf{Task.} Given a code review comment and a compact list of CWE-699 upper categories, determine whether the comment is \emph{security relevant}. If the comment is security relevant or potentially security relevant, select up to three relevant CWE upper categories and, when appropriate, identify candidate CWE subcategories.

  \vspace{0.25em}
  \textbf{Few-shot guidance.} Use annotated examples to calibrate decisions, especially for borderline or implicitly security-related comments:
  \emph{[Few-shot classification examples omitted for brevity]}

  \vspace{0.25em}
  \textbf{Decision principles.}
  \begin{itemize}[leftmargin=1.2em, itemsep=1pt, topsep=2pt, parsep=0pt]
    \item Prioritize recall: if any plausible security or safety concern exists, prefer \emph{uncertain} over \emph{no}.
    \item Do not require explicit vulnerability mentions; consider implicit risks such as memory safety, numeric errors, input validation, or privilege boundaries.
    \item Restrict category selection to the provided CWE-699 upper categories.
    \item Ground all decisions in concrete phrases drawn from the review comment.
  \end{itemize}

  \vspace{0.1em}
  \textbf{Output.} Produce a structured classification indicating whether the comment is security relevant, the selected CWE upper categories (if any), and brief signals justifying the decision.
  \end{promptbox}

  \caption{\textbf{System prompt used for Stage~1, Call~1}. Coarse-grained security-relevant review filtering in \eyeqllm{}.}
  \label{fig:stage1-call1-prompt}
\end{figure}

The second stage (see the corresponding prompt in \Cref{fig:stage1-call2-prompt}) performs fine-grained classification. Using only the CWE subcategories associated with the categories selected in the first stage, the model assigns a single, most specific CWE to the comment or rejects it as non-security-related. By dynamically slicing the taxonomy based on the first-stage output, this design reduces the context size by up to an order of magnitude while enabling definition-based matching rather than relying on keyword heuristics.


	\begin{figure}[t]
	  \centering
		  \begin{promptbox}
		  \textbf{System role.} You are a senior application security engineer performing final classification of code review comments.

  \vspace{0.25em}
  \textbf{Task.} Given (i) a code review comment and (ii) a filtered CWE context pack containing only candidate CWE categories/subcategories, determine whether the comment indicates a security weakness. If yes, select the \emph{single most specific} matching CWE subcategory from the provided context pack; otherwise, return \emph{security = no}.

  \vspace{0.25em}
  \textbf{Decision principles.}
  \begin{itemize}[leftmargin=1.2em, itemsep=1pt, topsep=2pt, parsep=0pt]
    \item Use \emph{only} the provided CWE context pack; do not use external knowledge or CWEs not present in the pack.
    \item Prefer \emph{definition match} (title + description) over keyword match.
    \item Choose the most specific applicable CWE subcategory in the pack.
    \item If the comment concerns readability, style, performance-only changes, or general refactoring with no plausible exploit/safety angle, output \emph{security = no}.
    \item When \emph{security = yes}, the selected CWE subcategory must appear in the provided context pack.
  \end{itemize}

  \vspace{0.15em}
  \textbf{Rationale.} Provide 1--3 sentences that cite concrete signals from the comment and connect them to the chosen CWE definition in the context pack.

  \vspace{0.15em}
  \textbf{Output.} Produce a structured result containing: (i) \emph{security} (\emph{yes/no}); (ii) the selected CWE subcategory (ID and title) if \emph{yes}, otherwise null; and (iii) the rationale.
  \end{promptbox}

  \caption{\textbf{System prompt used for Stage~1, Call~2}. Fine-grained CWE assignment using the filtered CWE context pack in \eyeqllm{}.}
  \label{fig:stage1-call2-prompt}
\end{figure}

To enhance the LLM’s ability, these prompts are designed following the few-shot learning paradigm, using examples collected from our human-based feasibility study of \eyeqhuman{}. Further details are discussed in the answer to RQ1 in the evaluation section. 

\subsection{Stage 2: Code Localization}

The input to the localization stage consists of one or more security-related review comments, the set of functions modified by the corresponding pull request, and, when available, a coarse security classification in the form of CWE subcategories derived from the previous stage. In this stage, the goal is to identify the functions related to each security comment, while leaving fine-grained, line-level localization to Stage~3. Localization is fundamentally a semantic matching problem between natural language and code, which makes it challenging in practice: reviewers often describe behavioral concerns rather than explicitly naming functions, pull requests may modify many functions simultaneously, and large codebases frequently rely on macros that obscure true function boundaries.

\subsubsection{Principles of localizing code} When reviewers explicitly mention code locations--such as function names or line numbers, which is relatively uncommon--this information can be used directly for localization. Otherwise, it is necessary to examine all relevant code changes and apply logical reasoning, grounded in an understanding of the codebase and its semantics, to identify the functions that correspond to the review comments. In the example below, we illustrate one such instance of this reasoning, highlighting the advantage of human intelligence in the localization process. For LLM, we can instill a similar capability through carefully designed few-shot examples.

\subsubsection{Example} Applying these principles to the motivating example, we manually examined the code changes associated with the pull request to identify the program logic responsible for enforcing fiber stack-size constraints. The review discussion centered on how stack exhaustion is prevented and how stack-size limits derived from user-provided configuration values are applied at runtime. This led us to \codeid{OnUpdateFiberStackSize} (in \path{Zend/zend.c}) as listed in Listing \ref{lst:fiber_overflow_c}, the handler that parses the \codeid{fiber.stack_size} INI configuration and commits the resulting value to the global fiber runtime state. This function represents the boundary at which untrusted configuration input is translated into concrete stack-size parameters that govern subsequent fiber execution. Its semantics therefore directly correspond to the reviewer's concerns about whether large or malformed stack-size values could violate stack-safety assumptions.

{
\begin{minted}[
  breaklines,
  breakanywhere,
  escapeinside=@@,
  xleftmargin=0.0em,
  xrightmargin=0.6em
]{c}
@\mylno{1}@// In Zend/zend.c: OnUpdateFiberStackSize()
@\mylno{2}@static ZEND_INI_MH(OnUpdateFiberStackSize)
@\mylno{3}@{
@\mylno{4}@    if (new_value) {
@\mylno{5}@        zend_long tmp = zend_ini_parse_quantity_warn(new_value, name);
@\mylno{6}@        if (tmp < 0) {
@\mylno{7}@            return FAILURE;
@\mylno{8}@        }
@\mylno{9}@        EG(fiber_stack_size) = tmp;
@\mylno{10}@    }
@\mylno{11}@    return SUCCESS;
@\mylno{12}@}
\end{minted}
\listingcaptionbelow{Fiber stack-size validation and state update in \capcode{OnUpdateFiberStackSize}}{lst:fiber_overflow_c}
}

\subsubsection{Automation with LLM} To automate this stage using an LLM in cases where functions are not explicitly mentioned, we decompose localization into two steps: candidate selection followed by code-diff verification. This design avoids a single monolithic prompt, which would exceed context limits for large pull requests. The process begins with a coarse candidate selection step, in which the LLM is asked to rank the most relevant functions among those modified in the pull request. The prompt used for this step (\Cref{fig:stage2-step1-prompt}) includes only function names and file paths, deliberately omitting code diffs to keep the context compact and scalable. The model is instructed to reason explicitly about the reviewer’s concern, grounding its ranking in textual or behavioral cues and prioritizing the function that owns the relevant invariant or implements the fix.

\begin{figure}[t]
  \centering
		  \begin{promptbox}
		  \textbf{System role.}
		  You are a security code review analyzer localizing reviewer concerns to functions modified in a pull request (PHP codebase).

  \vspace{0.25em}
  \textbf{Task.}
  Given a code review comment and a list of candidate functions modified in the pull request (file paths and function names only), select the most likely function(s) the comment refers to.  
  Return a short, ranked list. If no candidate can be justified, return an empty list.

  \vspace{0.25em}
  \textbf{Few-shot guidance.}
  Use annotated, project-specific localization examples to calibrate decisions, particularly when function ownership or invariants are implicit:

  \emph{[Few-shot localization examples omitted for brevity]}

  \vspace{0.25em}
  \textbf{Decision principles.}
  \begin{itemize}[leftmargin=1.2em, itemsep=1pt, topsep=2pt]
    \item Choose only from the provided candidates; do not invent functions.
    \item Prefer functions directly modified in the patch.
    \item Prefer the function that \emph{owns the invariant or implements the fix}, not merely where a bug manifests.
    \item Ground reasoning in concrete phrases or behavioral cues from the review comment.
    \item If no candidate can be confidently justified, abstain.
  \end{itemize}

  \vspace{0.2em}
  \textbf{Optional CWE hint (low weight).}
  When available, CWE category, subcategory, and rationale may be provided as auxiliary context to narrow the search space.  
  If the CWE hint conflicts with the review comment, the comment takes precedence.

  \vspace{0.2em}
  \textbf{Confidence.}
  Assign \emph{HIGH}, \emph{MEDIUM}, or \emph{LOW} confidence to distinguish strong matches from ambiguous cases.

  \vspace{0.2em}
  \textbf{Output.}
  Produce a structured result containing a one-line description of the concern, a ranked list of candidate functions with justification (including rejected alternatives), an overall confidence level, and whether CWE context was used.
  \end{promptbox}

  \caption{System prompt used for Stage~2, Step~1 candidate function selection using names and file paths in \eyeqllm{}.}
  \label{fig:stage2-step1-prompt}
\end{figure}

When a CWE classification is available from Stage 1, we inject it into this prompt as a low-weight hint. Specifically, the CWE category, subcategory, and accompanying rationale are provided as auxiliary context, with explicit instructions that the review comment takes precedence in the event of conflict. This design enables the model to use CWE information to narrow the search space (e.g., by favoring resource-management or initialization-related functions) while avoiding overfitting to potentially noisy classifications. The extended output schema records whether the CWE context was actually used, enabling downstream analysis and ablation studies.

Confidence levels are required to distinguish strong matches from ambiguous cases, and rejected candidates must be justified to discourage superficial reasoning. These constraints collectively steer the LLM toward precise, defensible localizations.

Candidate selection alone is insufficient, as function names can be misleading or may correspond to macros or prototypes rather than concrete definitions. We therefore introduce a second LLM-guided verification step that operates on the actual code diffs of the top-ranked candidates. The verification prompt (\Cref{fig:stage2-step2-prompt}) requires the model to align the review comment with specific code changes, explicitly citing lines or patterns from the diff that justify the localization. If none of the candidate diffs clearly matches the reviewer’s concern, the model is instructed to abstain rather than force a selection. This verification step substantially reduces hallucinated localizations and ensures that decisions are grounded in observable, source-level evidence rather than inferred behavior.

\begin{figure}[t]
  \centering
		  \begin{promptbox}
		  \textbf{System role.}
		  You are a security code review analyzer verifying candidate localizations using code diffs.

  \vspace{0.25em}
  \textbf{Task.}
  Given a review comment and diff snippets for the top-ranked candidate functions from Step~1, verify which function best matches the reviewer’s concern.  
  Select a function \emph{only if} the diff provides concrete supporting evidence.

  \vspace{0.25em}
  \textbf{Few-shot guidance.}
  Use annotated, project-specific verification examples to calibrate how review comments align with concrete code changes:

  \emph{[Few-shot localization examples omitted for brevity]}

  \vspace{0.25em}
  \textbf{Decision principles.}
  \begin{itemize}[leftmargin=1.2em, itemsep=1pt, topsep=2pt]
    \item Ground decisions strictly in \emph{actual code changes} shown in the diff.
    \item Cite the minimal diff line(s) that support the match.
    \item Prefer the function being fixed over callers or unrelated helpers.
    \item Do not infer behavior beyond what is visible in the diff.
    \item If no diff clearly matches the review comment, abstain rather than forcing a selection.
  \end{itemize}

  \vspace{0.2em}
  \textbf{Confidence.}
  Assign \emph{HIGH}, \emph{MEDIUM}, or \emph{LOW} confidence based on the strength of the diff evidence.

  \vspace{0.2em}
  \textbf{Output.}
  Produce a structured result containing the ranked candidate(s), cited diff evidence supporting the choice, rejected alternatives with justification, and an overall confidence level.
  \end{promptbox}

  \caption{System prompt used for Stage~2, Step~2 diff-based verification of localized functions in \eyeqllm{}.}
  \label{fig:stage2-step2-prompt}
\end{figure}

\subsection{Stage 3: Code Instrumentation}

Once a target function is identified in Stage 2 based on the review comments, the next step is to identify logic-relevant program elements (e.g., variables or expressions) and insert suitable IJON annotations to capture meaningful state changes. This stage is the most challenging because it requires fine-grained code comprehension to determine which elements influence the function’s control flow or key invariants highlighted in the review. It also requires selecting appropriate IJON annotation macros that expose these semantics as informative feedback signals during fuzzing.

\subsubsection{Principles of instrumenting code}

As IJON provides a variety of annotation macros, it is important to select those that align with the identified program elements and the review comments so the resulting signals guide exploration toward states of interest to reviewers. For example, \codeid{IJON_MIN} or \codeid{IJON_DIST} are suitable for detecting buffer-boundary issues, \codeid{IJON_BITS} and \codeid{IJON_MAX} work well for integer-related issues, and \codeid{IJON_STATE} and \codeid{IJON_CTX} are effective for capturing state-machine–related behaviors.

\subsubsection{Example} 

Inspection of \codeid{OnUpdateFiberStackSize} shows that the handler has a single semantic responsibility: it parses the user-provided INI string \codeid{new_value} (\hyperlink{lst-annotated_fiber_overflow_c-l4}{\Cref*{lst:annotated_fiber_overflow_c}, Line~4}), rejects invalid negative inputs (\hyperlink{lst-annotated_fiber_overflow_c-l5}{\Cref*{lst:annotated_fiber_overflow_c}, Lines~5--7}), and commits the resulting numeric value to \codeid{EG(fiber_stack_size)} (\hyperlink{lst-annotated_fiber_overflow_c-l8}{\Cref*{lst:annotated_fiber_overflow_c}, Line~8}). As reflected in the discussion in \Cref{fig:stack-overflow-fiber}, the developer concern is not whether this setter executes, but which concrete configuration
values are realized at runtime, as these values determine subsequent fiber execution behavior.

However, the control flow of the setter itself is largely invariant to the magnitude of the parsed value beyond basic validation, making coverage-based feedback insufficient for encouraging exploration of diverse configurations. We therefore annotate the committed stack-size value with \codeid{IJON_SET} (\hyperlink{lst-annotated_fiber_overflow_c-l11}{\Cref*{lst:annotated_fiber_overflow_c}, Line~11}), explicitly exposing \codeid{EG(fiber_stack_size)} as a fuzzing
state so that executions realizing distinct configuration values are prioritized.

\newpage
{
\begin{minted}[
  breaklines,
  breakanywhere,
  escapeinside=@@,
  xleftmargin=0.0em,
  xrightmargin=0.6em
]{c}
@\mylno{1}@static ZEND_INI_MH(OnUpdateFiberStackSize)
@\mylno{2}@{
@\mylno{3}@    if (new_value) {
@\mylno{4}@@\codeanchor{lst-annotated_fiber_overflow_c-l4}@        zend_long tmp = zend_ini_parse_quantity_warn(new_value, name);
@\mylno{5}@@\codeanchor{lst-annotated_fiber_overflow_c-l5}@        if (tmp < 0) {
@\mylno{6}@            return FAILURE;
@\mylno{7}@        }
@\mylno{8}@@\codeanchor{lst-annotated_fiber_overflow_c-l8}@        EG(fiber_stack_size) = tmp;
@\mylno{9}@
@\mylno{10}@        #ifdef _USE_IJON
@\mylno{11}@@\codeanchor{lst-annotated_fiber_overflow_c-l11}@        IJON_SET(EG(fiber_stack_size));  // TRACKS EXTREME VALUE
@\mylno{12}@        #endif
@\mylno{13}@    } else {
@\mylno{14}@        EG(fiber_stack_size) = ZEND_FIBER_DEFAULT_C_STACK_SIZE;
@\mylno{15}@    }
@\mylno{16}@    return SUCCESS;
@\mylno{17}@}
\end{minted}
\listingcaptionbelow{Adding IJON annotation to observe fiber stack-size updates in \capcode{OnUpdateFiberStackSize}}{lst:annotated_fiber_overflow_c}
}

\subsubsection{Automation with LLM}

We design a prompt (\Cref{fig:ijon-prompt-generate}) that first asks the LLM to understand the target function and identify candidate locations for annotations. Since LLMs are generally effective at code comprehension, particularly at the function level, this part of the prompt is relatively straightforward.

\begin{figure}[t]
\centering
\begin{promptbox}
\textbf{System role.}
You are an expert in AFL++ and IJON-guided fuzzing. Your task is to design high-quality IJON annotations for \emph{one} C/C++ function from the PHP codebase using its implementation and associated code-review signals.

\vspace{0.25em}
\textbf{Task.}
Given a function definition and related review comments, identify candidate locations within the function and propose IJON annotations that expose fragile invariants or security-relevant behaviors.
Propose the \emph{top five} annotation candidates (return fewer if fewer are justified).

\vspace{0.25em}
\textbf{Few-shot guidance.}
Use embedded, project-specific few-shot examples to calibrate macro selection, variable choice, and anchor placement:

\emph{[Few-shot IJON annotation examples omitted for brevity]}

\vspace{0.25em}
\textbf{IJON primitives.}
Select appropriate IJON macros from the available primitive set to expose security-relevant behaviors and fragile invariants, guided by the function logic, review comments, CWE context, and few-shot examples.

\emph{[Available IJON primitives omitted for brevity]}

\vspace{0.2em}
\textbf{Hard rules.}
\begin{itemize}[leftmargin=1.2em, itemsep=1pt, topsep=2pt]
\item Do not use line numbers; specify insertion points using anchors.
\item Each annotation must include a macro, compilable code snippet, insertion description, pre-anchor, post-anchor, and rationale.
\item Wrap injected code with \texttt{\#ifdef \_USE\_IJON} guards.
\item Do not introduce new headers or external symbols.
\item Prefer annotations aligned with reviewer concerns and fragile invariants.
\end{itemize}

\vspace{0.2em}
\textbf{Anchoring requirement.}
Anchors must be verbatim substrings occurring before and after the intended insertion location. This ensures robustness to formatting changes and minor patch drift while enabling precise placement.

\vspace{0.2em}
\textbf{Output.}
Produce a structured JSON object containing up to five IJON annotation candidates, each with anchors and a short rationale linking the annotation to review signals.
\end{promptbox}

\caption{System prompt used to generate anchored IJON annotation candidates for a single function in \eyeqllm{}.}
\label{fig:ijon-prompt-generate}
\end{figure}

Selecting appropriate IJON annotation macros, however, is more challenging. To address this, we apply several strategies. First, as shown in \Cref{fig:ijon-cwe-hint} we provide the CWE IDs selected in Stage~1, allowing the LLM to choose suitable macros based on the CWE descriptions, review comments, and few-shot examples. Second, we instruct the LLM to propose multiple annotations--up to five candidates--to increase the likelihood of covering relevant program behaviors.

\begin{figure}[t]
  \centering
		  \begin{promptbox}
		  \textbf{CWE classification hint.}
		  When available, the following CWE context is injected into the IJON prompt
		  as auxiliary guidance (low--medium weight):

  \begin{itemize}[leftmargin=1.2em, itemsep=1pt, topsep=2pt]
    \item Category: \texttt{<category\_title>} (CWE-\texttt{<category\_id>})
    \item Subcategory: \texttt{<subcategory\_title>} (CWE-\texttt{<subcategory\_id>})
    \item Rationale: \texttt{<classification rationale>}
  \end{itemize}

  \vspace{0.2em}
  \textbf{Usage guidance.}
  Use the CWE information only to inform macro selection and variable choice.
  If the CWE hint conflicts with the review comments, the comments take precedence.
  \end{promptbox}

  \caption{CWE classification injected as a soft prior to guide IJON annotation selection in \eyeqllm{}.}
  \label{fig:ijon-cwe-hint}
\end{figure}

Notably, the prompt asks the LLM to propose anchor points using verbatim substrings occurring before and after the intended insertion location, rather than line numbers. This approach makes annotation placement robust to formatting differences and minor patch drift, while still allowing precise insertion at the intended program point. 

\section{Implementation}
\label{sec:implementation}

\paragraph{Stage 1: Review classification.}
In Stage~1, we retrieve and aggregate review comments from GitHub pull requests, filtering and deduplicating them into a unified corpus. We then apply a two-stage LLM pipeline to classify comments for security relevance and assign CWE categories. We implemented this stage across six Python modules (4,334 LoC) for retrieval and 2,506 LoC for classification, totaling 6,840 LoC.

\paragraph{Stage 2: Code localization.}
In Stage~2, we map security-relevant review comments to concrete functions in the codebase using an LLM-assisted localization pipeline. We systematically verify candidate functions to ensure correctness. The Stage~2 implementation totals 10,220 LoC.

\paragraph{Stage 3: Code instrumentation.}
In Stage~3, we convert localized findings into executable fuzzing guidance by generating and injecting IJON annotations into the target functions. We perform safety checks to ensure syntactic correctness, variable-scope validity, and duplicate-free insertion. The instrumentation pipeline comprises 5,080 LoC.

\paragraph{Fuzzing infrastructure.}
We integrate the annotated code into a containerized AFL++-based fuzzing setup that supports parallel campaigns, crash replay, and post-fuzz triage. We also port IJON to AFL++ and modify it to use a dedicated bitmap for annotations, reducing collisions and improving fuzzing performance\footnote{\url{https://github.com/AFLplusplus/AFLplusplus/pull/2540}}\footnote{\url{https://github.com/AFLplusplus/AFLplusplus/pull/2542}}\footnote{\url{https://github.com/AFLplusplus/AFLplusplus/pull/2546}}. The changes to the fuzzer itself comprise 2,866 LoC, while the surrounding infrastructure--including scripts and Docker configurations--totals 6,537 LoC.

\section{Evaluation}
\label{sec:evaluation}

\noindent
We evaluate EyeQ's workflow along three dimensions: (1) whether a human analyst can follow the workflow to produce useful annotations, (2) whether an LLM can automate the workflow end to end, and (3) whether the automated system generalizes to newer, unseen review data.

\subsection{Research Questions}

To this end, we formulate three research questions.

\begin{itemize}
    \item \textbf{RQ1: Can a human analyst following the proposed workflow produce annotations that improve fuzzing effectiveness?} This research question evaluates the soundness of the proposed workflow itself. This serves as a baseline and demonstrates that the workflow is practically usable and effective.

    \item \textbf{RQ2: How effective is the LLM-based system in following the workflow and generating useful annotations?} This question evaluates the ability of \eyeqllm{} to automate the workflow. Using the same dataset as RQ1, we compare fuzzing outcomes and analyze the effectiveness of the system at different stages of the workflow.

    \item \textbf{RQ3: Does the LLM-based system generalize to unseen code reviews from the same project?} To avoid overfitting of \eyeqllm{} to the dataset used in RQ1 and RQ2, we evaluate it on a separate set of code reviews from the same software project. This question examines whether the system maintains effectiveness on previously unseen reviews.

\end{itemize}

\subsection{Experimental Setup}

For RQ1 and RQ2, we evaluate \eyeqhuman{} and \eyeqllm{}, respectively, using a dataset of labeled review comments from 2011 to 2022 drawn from a popular PHP project introduced by Charoenwet et al. \cite{charoenwet2024toward}. The dataset contains 240 review comments, evenly split between 120 security-related and 120 security-irrelevant comments. We randomly select three comments from each subset as few-shot examples for \eyeqllm{}, leaving 234 comments for evaluation. This dataset provides an established ground truth for the review classification stage of the workflow, which is critical for all downstream tasks.

To assess fuzzing effectiveness, we compare fuzzing outcomes on \eyeq{}-annotated code under both setups (\eyeqhuman{} and \eyeqllm{}). All experiments use IJON++, a recent IJON implementation that we ported to AFL++ and subsequently merged into the upstream AFL++ codebase, and are compared against standard AFL++ with IJON disabled.

For RQ1, our first author--a second-year PhD student with experience in fuzzing and program analysis but no prior familiarity with the PHP codebase--manually follows the workflow starting from Stage 2 (Code Localization) and Stage 3 (Code Instrumentation) to produce annotated source code for fuzzing with IJON. We omit Stage 1 (Review Classification) since the selected dataset already provides manually curated classification results. Stage 4 (Fuzz Driver Selection) is not evaluated because the main fuzz driver of the PHP project is used.

For RQ2, we evaluate \eyeqllm{} across all stages of the workflow, except Stage 4, which is also skipped as in RQ1. This includes Stage 1 (Review Classification), and we compare the results against the human-generated annotations and fuzzing outcomes from RQ1.

For RQ3, we evaluate \eyeqllm{} on recent artifacts from the PHP project collected between September 2025 and December 2025. This dataset includes 318 pull requests (PRs) and 965 review comments. We compare \eyeqllm{} \& IJON++ against AFL++ in terms of bug-finding effectiveness.

All experiments are conducted on identical virtual machines to ensure reproducibility and control for hardware-induced variability. Each VM is equipped with 32 virtual CPU cores, 128\,GB of RAM, and 1\,TB of local storage, running Ubuntu~22.04~LTS. We use the \texttt{gemini-2.5-flash} model with a fixed configuration across all experiments, limiting the maximum output length to 5{,}000 tokens to bound response verbosity and ensure consistent downstream parsing. No fine-tuning or task-specific adaptation is applied; all interactions rely solely on prompt engineering and the few-shot examples described earlier.

\subsection{RQ1: \texorpdfstring{\eyeqhuman{}}{EyeQ\_Human} vs AFL++}

From the 117 human-labeled security-related review comments in the selected dataset, the first author followed the \eyeq{} workflow to manually localize code and annotate functions. After the code localization stage, 37 reviews were mapped to 37 corresponding functions, resulting in a localization rate of 31.6\%. The remaining 80 reviews could not be mapped for two primary reasons. First, due to the size and complexity of the PHP codebase--with numerous submodules and intricate logic--the author was uncertain in many cases. This highlights the limitations of manual analysis and motivates the automated approach implemented in \eyeqllm{}. Second, several reviews referred to test files rather than source files. After the code instrumentation stage, the author successfully inserted IJON macros into all 37 functions, spanning 34 commits, and subsequently fuzzed them using IJON++. 

Running IJON++ on the 34 annotated code commits for 24 hours each uncovered 31 unique crashes (including the motivating example) after deduplication\footnote{Deduplication is based on crash call stacks and crash types, following common practice}. In contrast, AFL++, executed on the original code without annotations, found only 7 unique crashes, six of which were also discovered by IJON++ on the annotated code. Notably, AFL++ failed to detect the bug highlighted in the motivating example. 

It is worth noting that these unique crashes were counted based on the old commits in the review dataset. We were therefore interested in whether any of these crashes remain triggerable in the newest version of PHP at the time of our evaluation. We first re-ran the 31 crash-triggering inputs on the latest PHP release and observed that only 2 of them still persisted. However, this does not necessarily mean that the bugs associated with the remaining crashes have been fixed, since new features and code changes between the old and current versions could affect control and data flow, altering the originally crash-triggering inputs. To confirm this and further investigate the impact of our findings, we ported the annotations from the old versions to the newest PHP version and performed fuzzing with IJON++. With a single fuzzing instance running for four months, our experiments produced 41 previously unknown bugs; all were reported and confirmed by PHP project maintainers, with 27 of them fixed.

These bugs span a wide range of PHP subsystems, including hash table operations, object handling, fibers, garbage collection, configuration parsing, and the Zend virtual machine. Their concrete manifestations vary widely--from use‑after‑free and null‑pointer dereference to stack exhaustion and latent state‑consistency violations--highlighting diverse classes of interpreter errors that can lead to memory corruption or crashes. Detailed bug descriptions are provided in \Cref{tab:bug-finding-appx-all} in Appendix \ref{appx:bugs}.

This setup was used to assess whether the workflow remains tractable when multiple independent review-derived signals are present, not to suggest that aggregating annotations improves bug-finding effectiveness. The vulnerabilities discussed above are attributable to individual review signals, and fuzzing each localized region in isolation does not necessarily yield additional bugs. We combine annotations here for methodological reasons: to show that the workflow can accommodate multiple, independent review-guided annotations without interference, while preserving a clear correspondence between reviewer intent and the behaviors exercised during fuzzing.

  \begin{findingbox}{Finding 1}Humans can successfully follow the \eyeq{} workflow to extract insights from code reviews of the PHP codebase and translate them into annotations, enabling the discovery of 77\% more bugs compared to annotation-unaware fuzzing. However, humans face significant challenges in comprehending a large codebase and processing a large volume of code reviews, failing in 80 out of 117 cases (68.4\%). This limitation underscores the need for a more scalable, automated approach.
\end{findingbox}

\ifLuaTeX\vspace*{1.0em}\fi
\subsection{RQ2: \texorpdfstring{\eyeqllm{}}{EyeQ\_LLM} vs \texorpdfstring{\eyeqhuman{}}{EyeQ\_Human}}
\label{sec:evaluation-RQ1}

\ifLuaTeX\vspace{0.6em}\fi
\noindent
We next compare the LLM and human analysts across the full workflow and then drill down into per-stage agreements and disagreements.

\subsubsection{Per-Stage Comparison}

We first compare the performance of the LLM and humans at each stage of the \eyeq{} workflow. To ensure a fair and controlled comparison, we isolate each stage of the review-guided fuzzing process and apply both approaches to the same intermediate artifacts (234 original review comments, 117 security-related comments after Stage 1, and 37 human-localized functions after Stage 2). At each stage, we measure the agreement and divergence between \eyeqllm{} and the human judgments in \eyeqhuman{}. We present the results in the \eyeq{} workflow order, beginning with review classification and proceeding through localization, annotation selection and injection, and fuzzing.

\paragraph{LLM vs Human in Review Classification (Stage 1).}

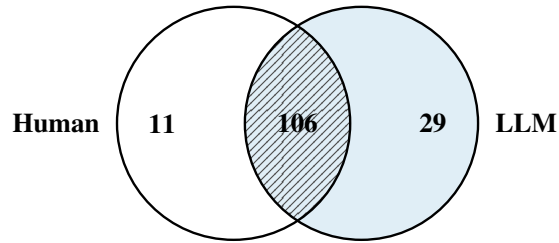
\begin{figure}[h]
  \centering
  \providecommand{\vennscale}{1}
\begin{tikzpicture}[font=\small, line cap=round, line join=round, scale=\vennscale]
  \path[use as bounding box] (-2.65,-1.75) rectangle (2.65,1.75);
  \definecolor{oiBlue}{HTML}{0072B2}
  \definecolor{oiOrange}{HTML}{E69F00}

  \def\r{1.55}
  \def\dx{0.85}
  \coordinate (H) at (-\dx, 0);
  \coordinate (L) at ( \dx, 0);

  \fill[oiBlue!12] (L) circle (\r);
  \begin{scope}
    \clip (H) circle (\r);
    \fill[oiOrange!18, pattern=north east lines, pattern color=black!70] (L) circle (\r);
  \end{scope}

  \draw[line width=0.9pt] (H) circle (\r);
  \draw[line width=0.9pt] (L) circle (\r);

  \node[font=\bfseries] at (0, 0) {106};
  \node[font=\bfseries] at (-1.80, 0) {11};
  \node[font=\bfseries] at ( 1.80, 0) {29};

  \node[font=\bfseries, anchor=east] at (-\dx-\r-0.10, 0) {Human};
  \node[font=\bfseries, anchor=west] at ( \dx+\r+0.10, 0) {LLM};

\end{tikzpicture}
  \caption{Agreements and disagreements between human analysts and the LLM in security-related comments.}
  \label{fig:stage1_venn}
\end{figure}

Of the 234 review comments in the dataset, the two approaches agree on 194 cases (82.9\%), including 106 comments that both identify as security-relevant. \Cref{fig:stage1_venn} also illustrates the remaining disagreements, which follow two patterns. 

In 29 cases, the LLM flags comments as security-relevant that humans label as non-security. These comments often contain technically suggestive terms such as overflow or type error, which \eyeqllm{} conservatively interprets as potential vulnerability indicators, even when the reviewer’s intent reflects a feature request or a general code-quality discussion. This behavior reflects a deliberate design choice that favors high recall, ensuring that potentially relevant signals are not prematurely discarded. Because \eyeqllm{} is fully automated and scalable, this design allows uncertain cases to be carried forward to subsequent stages.

LLM fails to identify 11 comments that humans classify as security-relevant. Manual inspection shows that these cases typically involve subtle or context-dependent security concerns, such as nuanced discussions of certificate validation semantics, where the security implication is implicit rather than explicitly framed in vulnerability-oriented language. In several instances, LLM recognizes relevant concepts but is constrained by classification requirements, leading to conservative rejection. We provide excerpts illustrating typical points of divergence between human analysts and LLM in \Cref{appx:classification} in Appendix \ref{appx:discussions}.

\paragraph{LLM vs Human in Code Localization (Stage 2).}

As reported in the answer to RQ1, the human (our first author) successfully localized code at the function level for only 41 out of 117 security-related comments. We randomly select 4 of these comments as few-shot examples for prompt construction and use the remaining 37 comments as ground truth to evaluate the LLM’s ability to localize code from review discussions. The results show that the human and the LLM agree on 18 review-comment–function pairs (48.6\%) and disagree in the remaining 19 cases.

Localization disagreements follow several patterns. First, the LLM may select a function that is conceptually related, such as choosing a verification routine instead of the corresponding creation routine. Second, when multiple functions implement nearly identical logic (e.g., encoding conversions with shared buffer-handling patterns), the LLM may select the wrong variant despite correctly identifying the underlying security concern. Third, when review comments explicitly discuss both a problematic construct and its replacement, the LLM may focus on the source of the issue rather than the function being introduced or modified. Examples of these disagreements can be found in \Cref{appx:localization} in Appendix~\ref{appx:discussions}.

\begin{figure*}[t]
  \centering
  \begingroup
  \captionsetup[subfigure]{font=small, labelfont=bf, justification=centering, singlelinecheck=true, skip=2pt}
  \newcommand{\venncard}[1]{%
    \begin{tcolorbox}[
      colback=black!2,
      colframe=black!20,
      boxrule=0.4pt,
      arc=2pt,
      left=3pt,right=3pt,top=0pt,bottom=11pt
    ]%
      \centering
      #1%
    \end{tcolorbox}%
  }
  \begin{tcolorbox}[
    enhanced,
    colback=blue!2!white,
    colframe=blue!35!black,
    colbacktitle=blue!10!white,
    coltitle=black,
    fonttitle=\small,
    boxrule=0.55pt,
    arc=4pt,
    drop shadow={black!20},
    left=6pt,right=6pt,top=6pt,bottom=4pt
  ]
  \begin{subfigure}[t]{0.32\linewidth}
    \centering
    \vspace{0pt}
    \venncard{{\def\vennscale{0.85}\providecommand{\vennscale}{1}
\begin{tikzpicture}[font=\small, line cap=round, line join=round, scale=\vennscale]
  \path[use as bounding box] (-2.65,-1.75) rectangle (2.65,1.75);

  \definecolor{oiBlue}{HTML}{0072B2}
  \definecolor{oiOrange}{HTML}{E69F00}
  \begin{scope}[xshift=-3mm, yshift=-2mm]
  \def\r{1.55}
  \def\dx{0.85}
  \coordinate (H) at (-\dx, 0);
  \coordinate (L) at ( \dx, 0);

  \fill[oiBlue!12] (L) circle (\r);
  \begin{scope}
    \clip (H) circle (\r);
    \fill[oiOrange!18, pattern=north east lines, pattern color=black!70] (L) circle (\r);
  \end{scope}

  \draw[line width=0.9pt] (H) circle (\r);
  \draw[line width=0.9pt] (L) circle (\r);

  \node[font=\bfseries] at (0, 0) {15};
  \node[font=\bfseries] at (-1.80, 0) {16};
  \node[font=\bfseries] at ( 1.80, 0) {9};

  \node[font=\bfseries, anchor=west] at (-\dx-\r+0.5, 0.95) {Human};
  \node[font=\bfseries, anchor=east] at ( \dx+\r-0.5, 0.95) {LLM};

  \end{scope}
\end{tikzpicture}}}
    \vspace{-0.4em}
    \subcaption{Stage~3 (isolated): human vs.\ LLM annotations (fixed localization).}
    \label{fig:bug_overlap_stage3_isolated}
  \end{subfigure}
  \hfill
  \begin{subfigure}[t]{0.32\linewidth}
    \centering
    \vspace{0pt}
    \venncard{{\def\vennscale{0.85}\providecommand{\vennscale}{1}
\begin{tikzpicture}[font=\small, line cap=round, line join=round, scale=\vennscale]
  \path[use as bounding box] (-2.65,-1.75) rectangle (2.65,1.75);

  \definecolor{oiBlue}{HTML}{0072B2}
  \definecolor{oiOrange}{HTML}{E69F00}
  \definecolor{oiGreen}{HTML}{009E73}
  \begin{scope}[xshift=-3mm, scale=0.92, transform shape]
  \coordinate (H) at (-1.25, 0.00);
  \coordinate (E) at ( 1.05, 0.00);
  \coordinate (A) at (-0.15,-0.85);
  \def\rH{1.35}
  \def\rE{1.55}
  \def\rA{0.85}

  \fill[oiOrange!25, fill opacity=0.35] (H) circle (\rH);
  \fill[oiBlue!20,   fill opacity=0.35] (E) circle (\rE);
  \fill[oiGreen!20,  fill opacity=0.35] (A) circle (\rA);

  \draw[line width=1.0pt] (H) circle (\rH);
  \draw[line width=1.0pt] (E) circle (\rE);
  \draw[line width=1.0pt] (A) circle (\rA);

  \node[font=\bfseries] at (-2.00, 0.20) {7};
  \node[font=\bfseries] at ( 2.00, 0.20) {23};
  \node[font=\bfseries] at ( -0.18, 0.35) {18};
  \node[font=\bfseries] at (-0.75,-0.85) {1};
  \node[font=\bfseries] at ( 0.45,-0.85) {1};
  \node[font=\bfseries] at (-0.15,-1.25) {0};
  \node[circle, fill=black!85, inner sep=2.2pt] at (-0.18,-0.45) {\textcolor{white}{\bfseries 5}};

  \node[font=\bfseries, anchor=west] at (-2.43, .72) {EyeQ\textsubscript{Human}};
  \node[font=\bfseries, anchor=east] at ( 1.8, .82) {EyeQ\textsubscript{LLM}};
  \node[font=\bfseries, anchor=north] at (-0,-1.62) {AFL++};

  \end{scope}
\end{tikzpicture}}}
    \vspace{-0.4em}
    \subcaption{End-to-end: human vs.\ LLM vs.\ AFL++ unique-crash overlap.}
    \label{fig:bug_overlap_end_to_end}
  \end{subfigure}
  \hfill
  \begin{subfigure}[t]{0.32\linewidth}
    \centering
    \vspace{0pt}
    \venncard{{\def\vennscale{0.85}\providecommand{\vennscale}{1}
\begin{tikzpicture}[font=\small, line cap=round, line join=round, scale=\vennscale]
  \path[use as bounding box] (-2.65,-1.75) rectangle (2.65,1.75);

  \definecolor{oiBlue}{HTML}{0072B2}
  \definecolor{oiOrange}{HTML}{E69F00}

  \begin{scope}[xshift=-3mm, yshift=-2mm]

  \def\r{1.55}
  \def\dx{0.85}
  \coordinate (H) at (-\dx, 0);
  \coordinate (L) at ( \dx, 0);

  \fill[oiBlue!12] (L) circle (\r);
  \begin{scope}
    \clip (H) circle (\r);
    \fill[oiOrange!18, pattern=north east lines, pattern color=black!70] (L) circle (\r);
  \end{scope}

  \draw[line width=0.9pt] (H) circle (\r);
  \draw[line width=0.9pt] (L) circle (\r);

  \node[font=\bfseries] at (0, 0) {9};
  \node[font=\bfseries] at (-1.80, 0) {21};
  \node[font=\bfseries] at ( 1.80, 0) {2};

  \node[font=\bfseries, anchor=west] at (-2.2, 0.80) {EyeQ\textsubscript{LLM}};
  \node[font=\bfseries, anchor=east] at ( 1.935, 1.05) {AFL++};

  \end{scope}
\end{tikzpicture}}}
    \vspace{-0.4em}
    \subcaption{PHP~2025: LLM vs.\ AFL++ unique-crash overlap.}
    \label{fig:bug_overlap_php2025}
  \end{subfigure}
  \end{tcolorbox}
  \caption{Overlap of unique crashes discovered under different fuzzing configurations.
 (a) \emph{Stage~3 (annotation-isolated):} crashes found using human-authored IJON annotations versus LLM-generated IJON annotations with fixed ground-truth localized functions, showing partial overlap and distinct crash sets.
 (b) \emph{End-to-end comparison:} overlap of unique crashes across \eyeqhuman{}, \eyeqllm{}, and baseline AFL++ when running the full EyeQ fuzzing pipeline.
 (c) \emph{RQ3 (PHP~2025):} direct comparison between \eyeqllm{} and AFL++ on a new, unseen PHP dataset (September–December 2025), showing a larger set of crashes unique to \eyeqllm{} with a smaller shared subset.}
  \label{fig:bug_overlap_venns}
  \endgroup
\end{figure*}
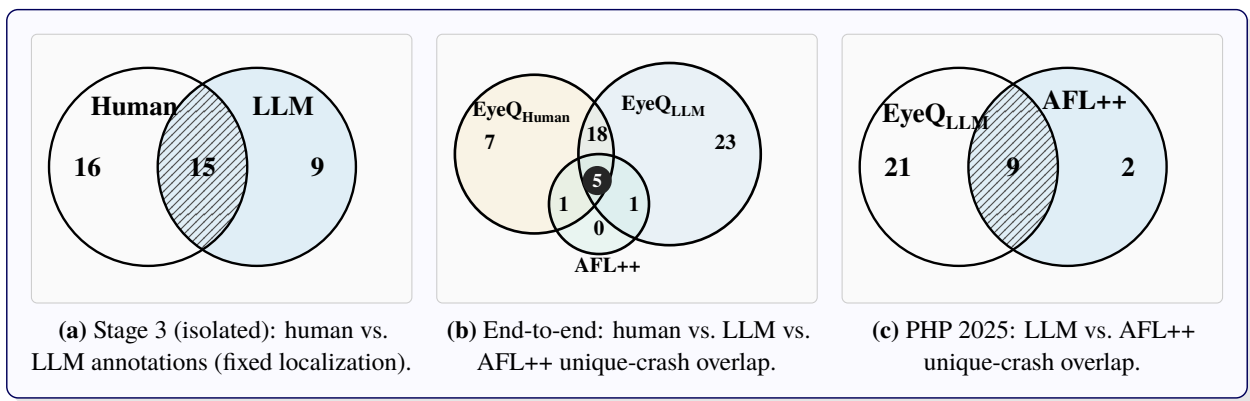

\paragraph{LLM vs Human in Code Instrumentation (Stage 3).}

To compare with human performance, we select the 37 functions identified by the human in Stage 2 and have the LLM automatically annotate them with IJON macros. Out of 37 evaluated cases, EyeQ produces annotations that align with human intent in 26 cases (70.3\%). 

At this stage, disagreements reflect differences in \emph{which semantic state is surfaced to the fuzzer}, rather than failures in identifying relevant code or review context. In agreement cases, LLM consistently exposes the same security-critical program state as human, including signature configuration flags, type states, nullability conditions, and resource lifecycle transitions. These annotations encode the same invariants and assumptions that human reviewers emphasize during security code review. Examples of these disagreements can be found in \Cref{appx:instrumentation} in Appendix~\ref{appx:discussions}.

Based on the annotations produced at this stage, we ran IJON++ to fuzz 34 commits for 24 hours each using the same setup as in RQ1. IJON++ discovered 24 bugs, of which 15 were also found in RQ1 using human-written annotations, while 9 were discovered only with LLM-generated annotations, as illustrated in Figure~\ref{fig:bug_overlap_venns}\subref{fig:bug_overlap_stage3_isolated}.

\begin{findingbox}{Finding 2} The per-stage comparison shows that the LLM’s performance is largely comparable to that of humans in review classification (Stage 1), achieving 82.9\% correctness, and in code instrumentation (Stage 3), with 70.3\% correctness. However, it performs substantially worse in Stage 2 (Code Localization), with only 48.6\% accuracy, which likely contributes to its weaker final fuzzing results (24 versus 31 bugs found).
\end{findingbox}

\subsubsection{End-to-End Comparison}

To complete RQ2, we evaluate \eyeqllm{} against \eyeqhuman{} in an end-to-end setting. Starting from the 234 review comments in the selected dataset, \eyeqllm{} automatically classifies the comments, identifying 135 as security-related, and successfully localizes code for 101 comments, achieving a substantially higher localization rate than humans (74.8\% vs. 31.6\%). These comments are mapped to 117 functions, into which \eyeqllm{} subsequently inserts 520 annotations. Finally, it fuzzes the resulting code across 99 commits for 24 hours each.

As shown in Figure~\ref{fig:bug_overlap_venns}\subref{fig:bug_overlap_end_to_end}, the results demonstrate that \eyeqllm{} significantly outperforms \eyeqhuman{}, discovering 47 bugs compared to 31. This improvement can be attributed to the substantially higher localization rate achieved by \eyeqllm{}, enabled by its scalable code comprehension capabilities, which allow more functions to be instrumented and fuzzed.

\begin{findingbox}{Finding 3}
Although the LLM performs worse than humans in terms of per-stage accuracy, \eyeqllm{} outperforms \eyeqhuman{} in the end-to-end setting due to its scalable code comprehension capabilities, which lead to higher localization rates and the insertion of more annotations.
\end{findingbox}

\subsection{RQ3: \texorpdfstring{\eyeqllm{}}{EyeQ\_LLM} Performance on a New Dataset}

In this research question, we evaluate the performance of \eyeqllm{} on a previously unseen dataset of recent PHP code reviews, without comparison to human performance. This experiment assesses whether \eyeqllm{} generalizes beyond the manually curated dataset used in RQ1 and RQ2 when applied to large-scale, real-world review data.

We focus on PHP code reviews from 2025, specifically from September to December. During this 4-month period, we collected 965 review comments from 318 pull requests.

\eyeqllm{} classified 445 of these comments as security-related and localized the corresponding functions. It then injected IJON annotations into the localized functions and aggregated them into 62 commits for fuzzing with IJON++, compared against AFL++.

The campaign uncovered 30 distinct crashes after deduplication. Among these, two correspond to previously unknown bugs in the PHP codebase at the time of discovery. The remaining crashes either overlap with bugs discovered in RQ1 or had been fixed previously. In comparison, AFL++, run under the same setup but without annotations, exposed only 11 unique crashes. Detailed results are summarized in \Cref{tab:rq3-fuzzing}.

\begin{findingbox}{Finding 4}
\eyeqllm{} is effective at finding bugs in an unseen PHP dataset and again outperforms annotation-unaware fuzzing by 2.7× (30 vs. 11).
\end{findingbox}

\section{Related Work}

Recent work on fuzzing has improved the scale and effectiveness of coverage-guided approaches through advances in mutation strategies, corpus management, and execution throughput \cite{pham2019smart, aschermann2019nautilus, aschermann2019redqueen,manes2019art,zhou2025kraken}. While these techniques make fuzzing practical at scale, they also reveal the limitations of conventional coverage signals in guiding exploration toward deep, semantically meaningful program states.

A complementary line of research focuses on richer feedback signals, including value- and state-aware instrumentation that captures progress beyond edge coverage \cite{gan2020greyone, fioraldi2021use,pham2020aflnet}. These approaches demonstrate that incorporating semantic cues into a fuzzer’s objectives can drive exploration into regions that would otherwise appear saturated under standard coverage metrics.

Human-in-the-loop fuzzing \cite{shoshitaishvili2017rise, fang2024ddgf, gao2025insightql, aschermann2020ijon} highlights the value of developer insight, showing that manual guidance--through grammars, rules, or interactive steering--can improve exploration. Prior techniques, however, typically require explicit human effort during fuzzing. EyeQ differs by extracting guidance from artifacts already produced during normal development, avoiding additional manual effort during fuzzing runs.

Directed fuzzing aims to prioritize inputs that exercise specific code regions or vulnerabilities, improving efficiency over traditional coverage-guided approaches. Popular directed fuzzers, such as AFLGo \cite{bohme2017directed}, guide mutations toward user-specified targets using static analysis and distance metrics. \eyeq{} differs by indirectly directing exploration through annotations: annotated functions are preferentially fuzzed, increasing the likelihood that test inputs exercise these security-relevant areas.

Finally, research on code reviews has shown that review discussions encode valuable reasoning about correctness and risk, even when not explicitly framed as vulnerability reports \cite{charoenwet2024toward}. \eyeq{} leverages this insight by transforming review signals into concrete instrumentation, bridging review-time knowledge with test-time fuzzing to improve bug-finding effectiveness.

\section{Conclusions}
This paper presents EyeQ, a workflow that transforms code review discussions into annotation-aware fuzzing guidance. We view it as an opening gambit in leveraging developer insights.
Our results are encouraging but not a panacea---the annotations appear to improve quality of fuzzing through an improved corpus without requiring substantial human effort. Yet, we do not yet actively \textit{direct} the fuzzer toward the program states implicated by these annotations, as done by directed fuzzing approaches such as AFLGo~\cite{bohme2017directed}. This leaves ample room for improvement. Nonetheless it is our view that the work is instructional and provides stable ground on which to begin a more concerted chase of code's dragons \dragon.

\section*{Ethical Considerations}
This work analyzes only publicly available code-review artifacts and source-code changes. When our experiments surfaced potential security-relevant issues, we followed responsible disclosure practices by notifying the appropriate project maintainers privately (or via the project's preferred security channel), sharing minimal proof-of-concept details needed to reproduce the issue, and allowing reasonable time for triage and fixes before any public discussion. Our fuzzing and instrumentation were only performed in controlled environments to avoid impacting production systems.

\section*{Open Science}

We will open-source the accompanying code upon acceptance of the paper to facilitate further research in this direction.

\section*{Acknowledgement}

Shaanan Cohney and Van-Thuan Pham were supported by two Australian Research Council’s Discovery Early Career Researcher Award (DECRA) projects (DE260100249 and DE230100473). We thank Google Cloud for providing research  credits to partially conduct our experiments.

\printbibliography

@String{Computing = "Computing" }

@String{Computer = "{IEEE} Computer" }

@String{Springer = "Springer-Verlag" }

@article{charoenwet2024toward,
  title={Toward effective secure code reviews: an empirical study of security-related coding weaknesses},
  author={Charoenwet, Wachiraphan and Thongtanunam, Patanamon and Pham, Van-Thuan and Treude, Christoph},
  journal={Empirical Software Engineering},
  volume={29},
  number={4},
  pages={88},
  year={2024},
  publisher={Springer}
}

@inproceedings{aschermann2020ijon,
  title={Ijon: Exploring deep state spaces via fuzzing},
  author={Aschermann, Cornelius and Schumilo, Sergej and Abbasi, Ali and Holz, Thorsten},
  booktitle={2020 IEEE Symposium on Security and Privacy (SP)},
  pages={1597--1612},
  year={2020},
  organization={IEEE}
}

@misc{ossfuzz_blogpost,
  author       = {Jonathan Metzman and Abhishek Arya and Oliver Chang and Kostya Serebryany},
  title        = {OSS-Fuzz --- Continuous Fuzzing for Open Source Software},
  howpublished = {\url{https://opensource.googleblog.com/2017/05/oss-fuzz-five-months-later-and.html}},
  year         = {2017},
  note         = {Google Open Source Blog, Accessed: 2025-10-10}
}

@inproceedings{bacchelli2013expectations,
  title={Expectations, outcomes, and challenges of modern code review},
  author={Bacchelli, Alberto and Bird, Christian},
  booktitle={2013 35th International Conference on Software Engineering (ICSE)},
  pages={712--721},
  year={2013},
  organization={IEEE}
}

@incollection{fagan2011design,
  title={Design and code inspections to reduce errors in program development},
  author={Fagan, Michael},
  booktitle={Software pioneers: contributions to software engineering},
  pages={575--607},
  year={2011},
  publisher={Springer}
}

@inproceedings{rigby2013convergent,
  title={Convergent contemporary software peer review practices},
  author={Rigby, Peter C and Bird, Christian},
  booktitle={Proceedings of the 2013 9th joint meeting on foundations of software engineering},
  pages={202--212},
  year={2013}
}

@inproceedings{balachandran2013reducing,
  title={Reducing human effort and improving quality in peer code reviews using automatic static analysis and reviewer recommendation},
  author={Balachandran, Vipin},
  booktitle={2013 35th International Conference on Software Engineering (ICSE)},
  pages={931--940},
  year={2013},
  organization={IEEE}
}

@article{mantyla2008types,
  title={What types of defects are really discovered in code reviews?},
  author={M{\"a}ntyl{\"a}, Mika V and Lassenius, Casper},
  journal={IEEE Transactions on Software Engineering},
  volume={35},
  number={3},
  pages={430--448},
  year={2008},
  publisher={IEEE}
}

@inproceedings{fioraldi2020afl++,
  title={$\{$AFL++$\}$: Combining incremental steps of fuzzing research},
  author={Fioraldi, Andrea and Maier, Dominik and Ei{\ss}feldt, Heiko and Heuse, Marc},
  booktitle={14th USENIX workshop on offensive technologies (WOOT 20)},
  year={2020}
}

@inproceedings{fioraldi2022libafl,
  title={Libafl: A framework to build modular and reusable fuzzers},
  author={Fioraldi, Andrea and Maier, Dominik Christian and Zhang, Dongjia and Balzarotti, Davide},
  booktitle={Proceedings of the 2022 ACM SIGSAC Conference on Computer and Communications Security},
  pages={1051--1065},
  year={2022}
}

@inproceedings{charoenwet2024empirical,
  title={An empirical study of static analysis tools for secure code review},
  author={Charoenwet, Wachiraphan and Thongtanunam, Patanamon and Pham, Van-Thuan and Treude, Christoph},
  booktitle={Proceedings of the 33rd ACM SIGSOFT international symposium on software testing and analysis},
  pages={691--703},
  year={2024}
}

@inproceedings{aschermann2019redqueen,
  title={REDQUEEN: Fuzzing with Input-to-State Correspondence.},
  author={Aschermann, Cornelius and Schumilo, Sergej and Blazytko, Tim and Gawlik, Robert and Holz, Thorsten},
  booktitle={NDSS},
  volume={19},
  pages={1--15},
  year={2019}
}

@inproceedings{pham2020aflnet,
  title={Aflnet: A greybox fuzzer for network protocols},
  author={Pham, Van-Thuan and B{\"o}hme, Marcel and Roychoudhury, Abhik},
  booktitle={2020 IEEE 13th International Conference on Software Testing, Validation and Verification (ICST)},
  pages={460--465},
  year={2020},
  organization={IEEE}
}

@article{baldoni2018survey,
  title={A survey of symbolic execution techniques},
  author={Baldoni, Roberto and Coppa, Emilio and D'elia, Daniele Cono and Demetrescu, Camil and Finocchi, Irene},
  journal={ACM Computing Surveys (CSUR)},
  volume={51},
  number={3},
  pages={1--39},
  year={2018},
  publisher={ACM New York, NY, USA}
}

@inproceedings{gao2023beyond,
  title={Beyond the coverage plateau: A comprehensive study of fuzz blockers (registered report)},
  author={Gao, Wentao and Pham, Van-Thuan and Liu, Dongge and Chang, Oliver and Murray, Toby and Rubinstein, Benjamin IP},
  booktitle={Proceedings of the 2nd International Fuzzing Workshop},
  pages={47--55},
  year={2023}
}

@inproceedings{babic2019fudge,
  title={Fudge: fuzz driver generation at scale},
  author={Babi{\'c}, Domagoj and Bucur, Stefan and Chen, Yaohui and Ivan{\v{c}}i{\'c}, Franjo and King, Tim and Kusano, Markus and Lemieux, Caroline and Szekeres, L{\'a}szl{\'o} and Wang, Wei},
  booktitle={Proceedings of the 2019 27th ACM Joint Meeting on European Software Engineering Conference and Symposium on the Foundations of Software Engineering},
  pages={975--985},
  year={2019}
}

@inproceedings{ispoglou2020fuzzgen,
  title={$\{$FuzzGen$\}$: Automatic fuzzer generation},
  author={Ispoglou, Kyriakos and Austin, Daniel and Mohan, Vishwath and Payer, Mathias},
  booktitle={29th USENIX Security Symposium (USENIX Security 20)},
  pages={2271--2287},
  year={2020}
}

@inproceedings{chen2023hopper,
  title={Hopper: Interpretative fuzzing for libraries},
  author={Chen, Peng and Xie, Yuxuan and Lyu, Yunlong and Wang, Yuxiao and Chen, Hao},
  booktitle={Proceedings of the 2023 ACM SIGSAC Conference on Computer and Communications Security},
  pages={1600--1614},
  year={2023}
}

@article{pham2019smart,
  title={Smart greybox fuzzing},
  author={Pham, Van-Thuan and B{\"o}hme, Marcel and Santosa, Andrew E and C{\u{a}}ciulescu, Alexandru R{\u{a}}zvan and Roychoudhury, Abhik},
  journal={IEEE Transactions on Software Engineering},
  volume={47},
  number={9},
  pages={1980--1997},
  year={2019},
  publisher={IEEE}
}

@inproceedings{aschermann2019nautilus,
  title={NAUTILUS: Fishing for deep bugs with grammars.},
  author={Aschermann, Cornelius and Frassetto, Tommaso and Holz, Thorsten and Jauernig, Patrick and Sadeghi, Ahmad-Reza and Teuchert, Daniel},
  booktitle={NDSS},
  volume={19},
  pages={337},
  year={2019}
}

@inproceedings{fioraldi2021use,
  title={The use of likely invariants as feedback for fuzzers},
  author={Fioraldi, Andrea and D'Elia, Daniele Cono and Balzarotti, Davide},
  booktitle={30th USENIX Security Symposium (USENIX Security 21)},
  pages={2829--2846},
  year={2021}
}

@inproceedings{gan2020greyone,
  title={$\{$GREYONE$\}$: Data flow sensitive fuzzing},
  author={Gan, Shuitao and Zhang, Chao and Chen, Peng and Zhao, Bodong and Qin, Xiaojun and Wu, Dong and Chen, Zuoning},
  booktitle={29th USENIX security symposium (USENIX Security 20)},
  pages={2577--2594},
  year={2020}
}

@inproceedings{shoshitaishvili2017rise,
  title={Rise of the hacrs: Augmenting autonomous cyber reasoning systems with human assistance},
  author={Shoshitaishvili, Yan and Weissbacher, Michael and Dresel, Lukas and Salls, Christopher and Wang, Ruoyu and Kruegel, Christopher and Vigna, Giovanni},
  booktitle={Proceedings of the 2017 ACM SIGSAC Conference on Computer and Communications Security},
  pages={347--362},
  year={2017}
}

@article{gao2025insightql,
  title={InsightQL: Advancing Human-Assisted Fuzzing with a Unified Code Database and Parameterized Query Interface},
  author={Gao, Wentao and Borovica-Gajic, Renata and Cha, Sang Kil and Qiu, Tian and Pham, Van-Thuan},
  journal={arXiv preprint arXiv:2510.04835},
  year={2025}
}

@inproceedings{fang2024ddgf,
  title={DDGF: Dynamic Directed Greybox Fuzzing with Path Profiling},
  author={Fang, Haoran and Zhang, Kaikai and Yu, Donghui and Zhang, Yuanyuan},
  booktitle={Proceedings of the 33rd ACM SIGSOFT International Symposium on Software Testing and Analysis},
  pages={832--843},
  year={2024}
}

@inproceedings{she2022effective,
  title={Effective seed scheduling for fuzzing with graph centrality analysis},
  author={She, Dongdong and Shah, Abhishek and Jana, Suman},
  booktitle={2022 IEEE Symposium on Security and Privacy (SP)},
  pages={2194--2211},
  year={2022},
  organization={IEEE}
}

@inproceedings{bohme2020boosting,
  title={Boosting fuzzer efficiency: An information theoretic perspective},
  author={B{\"o}hme, Marcel and Man{\`e}s, Valentin JM and Cha, Sang Kil},
  booktitle={Proceedings of the 28th ACM Joint Meeting on European Software Engineering Conference and Symposium on the Foundations of Software Engineering},
  pages={678--689},
  year={2020}
}

@article{manes2019art,
  title={The art, science, and engineering of fuzzing: A survey},
  author={Man{\`e}s, Valentin JM and Han, HyungSeok and Han, Choongwoo and Cha, Sang Kil and Egele, Manuel and Schwartz, Edward J and Woo, Maverick},
  journal={IEEE Transactions on Software Engineering},
  volume={47},
  number={11},
  pages={2312--2331},
  year={2019},
  publisher={IEEE}
}

@inproceedings{bohme2017directed,
  title={Directed greybox fuzzing},
  author={B{\"o}hme, Marcel and Pham, Van-Thuan and Nguyen, Manh-Dung and Roychoudhury, Abhik},
  booktitle={Proceedings of the 2017 ACM SIGSAC conference on computer and communications security},
  pages={2329--2344},
  year={2017}
}

@article{zhou2025kraken,
  title={KRAKEN: Program-Adaptive Parallel Fuzzing},
  author={Zhou, Anshunkang and Huang, Heqing and Zhang, Charles},
  journal={Proceedings of the ACM on Software Engineering},
  volume={2},
  number={ISSTA},
  pages={274--296},
  year={2025},
  publisher={ACM New York, NY, USA}
}

\appendix
\section{Detailed Discussion} \label{appx:discussions}
We provide some additional detail that may be of interest and illustrative to those seeking a deeper understanding of the interaction between code review commentary and the relevant code (and bugs).

\subsection{Classification} \label{appx:classification}
The following excerpts illustrate typical points of divergence between human analysts and EyeQ:



\begin{reviewquotebox}
\texttt{``It would be great if we could warn on overflow and on truncation to 32-bit.
Detecting overflow is pretty simple.''}
\end{reviewquotebox}


\begin{reviewquotebox}
\texttt{``It seems to be a type error for me.''}
\end{reviewquotebox}

In both cases above, EyeQ classifies the comment as security-relevant due to the presence of vulnerability-associated terminology, while human analysts interpret the discussion as non-security feature requests or general observations.


\begin{reviewquotebox}
\texttt{``Two general notes: I don't think we should modify the signature of
set\_ssl\ldots Probably we should land something similar to the patch proposed\ldots
to allow disabling certificate validation\ldots Does this completely turn off
verification of the host cert, or does it just disable checking the CN?''}
\end{reviewquotebox}

In this case, the human analyst identifies a clear security concern related to SSL/TLS certificate verification semantics, while EyeQ fails to classify the comment as security-relevant due to limitations imposed by its classification constraints. Taken together, these examples show that EyeQ's review signal extraction closely mirrors human judgment at the point where developers decide whether a discussion warrants security attention.

\subsection{Localization} \label{appx:localization}

The following example illustrates a representative case of semantically close but incorrect localization:

\begin{reviewquotebox}
\texttt{``I agree that this change makes sense, but the internal hashes do not provide any security, since anybody who is able to tamper with the phar can easily modify the hashes as well -- the internal hash only shields against accidental corruption of the phar. Only an external hash can provide security\ldots Properly securing phars would require something like \texttt{phar.require\_signature}, i.e. enforcing the phar to have an OpenSSL signature; and even then proper signature verification would be required.''}
\end{reviewquotebox}

In this case, LLM correctly identifies the discussion as concerning phar signature security and selects \texttt{phar\_\allowbreak verify\_signature}, while the human ground truth localizes the change to \texttt{phar\_create\_\allowbreak signature}. Both functions operate within the same security-critical subsystem, and the review comment itself discusses both creation and verification semantics, making the distinction inherently ambiguous.

\subsection{Code Instrumentation} \label{appx:instrumentation}

\paragraph{Example: Signature Security (Semantic Match)}
The following review explicitly argues that internal hashes do not provide security guarantees and that cryptographic signature enforcement is required:

\begin{reviewquotebox}
\texttt{``I agree that this change makes sense, but the internal hashes do not provide any security, since anybody who is able to tamper with the phar can easily modify the hashes as well -- the internal hash only shields against accidental corruption of the phar. Only an external hash can provide security\ldots Properly securing phars would require something like \texttt{phar.require\_signature}, i.e. enforcing the phar to have an OpenSSL signature; and even then proper signature verification would be required.''}
\end{reviewquotebox}

\Cref{lst:phar_sig_human} shows the human written annotation, which tracks the signature configuration flag \texttt{phar\allowbreak->sig\_\allowbreak flags} at points where the signature algorithm is selected. \Cref{lst:phar_sig_llm} shows the LLM-generated annotation, which instruments the same variable inside the signature-generation logic. Although the LLM inserts annotations at finer granularity and across multiple control-flow paths, both annotations expose the same security-critical state governing cryptographic strength.

\vspace{0.4em}
\noindent
\begin{minipage}{\dimexpr\linewidth-12pt\relax}
\begin{minted}[
  breaklines,
  breakanywhere,
  escapeinside=@@,
  xleftmargin=0.0em,
  xrightmargin=0.0em
]{c}
// Function: phar_create_signature
@\mylno{1}@default:
@\mylno{2}@    #ifdef _USE_IJON
@\mylno{3}@    IJON_SET(phar->sig_flags);
@\mylno{4}@    #endif
@\mylno{5}@    phar->sig_flags = PHAR_SIG_SHA256;
@\mylno{6}@    ZEND_FALLTHROUGH;

@\mylno{7}@if (!phar->is_data && !phar->sig_flags) {
@\mylno{8}@    #ifdef _USE_IJON
@\mylno{9}@    IJON_SET(phar->sig_flags);
@\mylno{10}@    #endif
@\mylno{11}@    phar->sig_flags = PHAR_SIG_SHA256;
@\mylno{12}@}
\end{minted}
\listingcaptionbelow{Human IJON Annotation for Phar Signature Configuration}{lst:phar_sig_human}
\end{minipage}

\vspace{0.55em}
\noindent
\begin{minipage}{\dimexpr\linewidth-12pt\relax}
\begin{minted}[
  breaklines,
  breakanywhere,
  escapeinside=@@,
  xleftmargin=0.0em,
  xrightmargin=0.0em
]{c}
@\mylno{1}@int phar_create_signature(phar_archive_data *phar, php_stream *fp, ...)
@\mylno{2}@{
@\mylno{3}@#ifdef _USE_IJON
@\mylno{4}@    IJON_STATE(1);
@\mylno{5}@#endif
@\mylno{6}@    switch (phar->sig_flags) {
@\mylno{7}@#ifdef _USE_IJON
@\mylno{8}@        IJON_SET(phar->sig_flags);
@\mylno{9}@#endif
@\mylno{10}@        case PHAR_SIG_SHA256:
@\mylno{11}@            ...
@\mylno{12}@        case PHAR_SIG_OPENSSL:
@\mylno{13}@#ifdef _USE_IJON
@\mylno{14}@            IJON_MAX(sig_len);
@\mylno{15}@            IJON_SET(siglen);
@\mylno{16}@#endif
@\mylno{17}@            ...
@\mylno{18}@    }
@\mylno{19}@    return SUCCESS;
@\mylno{20}@}
\end{minted}
\listingcaptionbelow{LLM-Generated IJON Annotation for Phar Signature Configuration}{lst:phar_sig_llm}
\end{minipage}

\paragraph{Example: Null-Pointer Guard (Semantic Divergence).}
The following review points out that an existing null check is ineffective because the dereference occurs earlier:


\begin{reviewquotebox}
\texttt{``Then should the if(!pReq) code be removed? Because a null pointer dereference will occur before it reaches that.''}
\end{reviewquotebox}

\Cref{lst:null_guard_human} shows the human annotation, which encodes a defensive invariant by tracking whether execution passes through the null-check guard before pointer use. In contrast, \Cref{lst:null_guard_llm} shows the LLM-generated annotation, which instruments the pointer value prior to the guard, thereby exposing the vulnerable dereference rather than the intended safety condition.

\par\smallskip\noindent
\begin{minipage}{\dimexpr\linewidth-12pt\relax}
\begin{minted}[
  breaklines,
  breakanywhere,
  escapeinside=@@,
  xleftmargin=0.0em,
  xrightmargin=0.0em
]{c}
// Function: LSAPI_GetEnv_r
@\mylno{1}@if (!pReq || !name) {
@\mylno{2}@    #ifdef _USE_IJON
@\mylno{3}@    IJON_INC(!pReq || !name);
@\mylno{4}@    #endif
@\mylno{5}@    return NULL;
@\mylno{6}@}
@\mylno{7}@#ifdef _USE_IJON
@\mylno{8}@IJON_INC(pReq && name);
@\mylno{9}@#endif
\end{minted}
\listingcaptionbelow{Human IJON Annotation Encoding a Null-Check Invariant}{lst:null_guard_human}
\end{minipage}

\Needspace{18\baselineskip}
\par\smallskip\noindent
\begin{minipage}{\dimexpr\linewidth-12pt\relax}
\begin{minted}[
  breaklines,
  breakanywhere,
  escapeinside=@@,
  xleftmargin=0.0em,
  xrightmargin=0.0em
]{c}
@\mylno{1}@char *LSAPI_GetEnv_r(LSAPI_Request *pReq, const char *name)
@\mylno{2}@{
@\mylno{3}@#ifdef _USE_IJON
@\mylno{4}@    IJON_SET((uintptr_t)pReq);
@\mylno{5}@#endif
@\mylno{6}@    struct LSAPI_key_value_pair *pBegin =
@\mylno{7}@        pReq->m_pEnvList;  // Dereference before guard
@\mylno{8}@    ...
@\mylno{9}@    if (!pReq || !name)
@\mylno{10}@        return NULL;
@\mylno{11}@}
\end{minted}
\listingcaptionbelow{LLM-Generated IJON Annotation Demonstrating Premature Dereference}{lst:null_guard_llm}
\end{minipage}

Taken together, these results show that EyeQ frequently captures the same semantic vulnerability states as human experts when generating IJON annotations, while divergences arise primarily when reviewer intent requires encoding a defensive invariant rather than exposing an immediately observable program state.

\section{Bug finding results}
\label{appx:bugs}

This appendix provides the detailed crash and bug lists referenced throughout the evaluation. We first list the unique crashes from RQ2 and RQ3, then summarize the previously unknown bugs found in our longer-running study.

\providecommand{\codeid}[1]{\texttt{#1}}
\newcommand{\eyeqapptablesize}{\scriptsize}
\newcommand{\eyeqapptablearraystretch}{1.08}
\newcommand{\eyeqapptableemergencystretch}{2em}
\newenvironment{eyeqapptable}[1]{%
  \begingroup
  \eyeqapptablesize
  \renewcommand{\arraystretch}{\eyeqapptablearraystretch}%
  \setlength{\tabcolsep}{#1}%
  \setlength{\emergencystretch}{\eyeqapptableemergencystretch}%
  \setlength\LTleft{0pt}%
  \setlength\LTright{0pt}%
  \captionsetup{justification=centering,singlelinecheck=true,hypcap=false}%
}{%
  \endgroup
}

\subsection{Unique crashes in RQ2}

\Cref{tab:rq2-fuzzing-dedup} lists all unique crashes found by \eyeqllm{} in the experiments for RQ2.

\begin{eyeqapptable}{3pt}
\begin{longtable}{
>{\centering\arraybackslash}p{0.65cm}
>{\RaggedRight\arraybackslash}p{3.0cm}
>{\RaggedRight\arraybackslash}p{3.4cm}
>{\centering\arraybackslash}p{0.75cm}
>{\RaggedRight\arraybackslash}p{3.6cm}
>{\centering\arraybackslash}p{1.1cm}
>{\RaggedRight\arraybackslash}p{\dimexpr\textwidth-12.5cm-14\tabcolsep\relax}
}
\caption{RQ2 --- 24 unique crashes found by \eyeqllm{} in the per-stage comparison.}%
\label{tab:rq2-fuzzing-dedup}\\
\toprule
\textbf{ID} & \textbf{Function} & \textbf{Source File} & \textbf{Occ.} & \textbf{Commits} & \textbf{PHP Ver.} & \textbf{ASAN Summary} \\
\midrule
\endfirsthead
\toprule
\textbf{ID} & \textbf{Function} & \textbf{Source File} & \textbf{Occ.} & \textbf{Commits} & \textbf{PHP Ver.} & \textbf{ASAN Summary} \\
\midrule
\endhead
\bottomrule
\endfoot
\bottomrule
\endlastfoot

01 & \path{php_pcre_create_match_data} &
\path{ext/pcre/php_pcre.c} &
242 &
\makecell[l]{\texttt{57a4a2c5a8dd}\\\texttt{7553c696c306}\\\texttt{2aceb0b00e21}\\\texttt{312fa05a291b}\\\texttt{df2414a2230d}\\\texttt{182e3ac0c399}\\\texttt{28500fe4ef12}\\\texttt{e7123ef569d2}\\\texttt{11bf5ccf6238}\\\texttt{8bb7417ca26c}\\\texttt{bebb0b7bbe0c}\\\texttt{c945c399aa80}\\\texttt{d54e197baee6}} &
\makecell[c]{8.0\\8.1\\8.2} &
ABRT in \texttt{raise}. \\

02 & \path{zval_undefined_cv} &
\path{Zend/zend_execute.c} &
3 &
\makecell[l]{\texttt{7553c696c306}\\\texttt{bebb0b7bbe0c}} &
\makecell[c]{8.0\\8.1} &
SEGV in \path{zval_undefined_cv}. \\

03 & \path{zend_gc_delref} &
\path{Zend/zend_types.h} &
5 &
\makecell[l]{\texttt{11bf5ccf6238}\\\texttt{8bb7417ca26c}\\\texttt{d54e197baee6}} &
\makecell[c]{8.1} &
SEGV in GC reference handling. \\

04 & \path{decrement_function} &
\path{Zend/zend_operators.c} &
13 &
\makecell[l]{\texttt{57a4a2c5a8dd}} &
\makecell[c]{8.0} &
ABRT in \texttt{raise}. \\

05 & \path{__interceptor_strlen} &
\path{main/spprintf.c} &
42 &
\makecell[l]{\texttt{3fe698b90493}\\\texttt{e23a9d6a5c7d}\\\texttt{03108697b86f}\\\texttt{08755e6b8915}\\\texttt{2d48d734a201}\\\texttt{c945c399aa80}} &
\makecell[c]{7.3\\7.4\\8.1} &
Stack overflow in \texttt{strlen}. \\

06 & \path{__interceptor_strlen} &
\path{main/snprintf.c} &
1 &
\makecell[l]{\texttt{d54e197baee6}} &
\makecell[c]{8.1} &
SEGV in \texttt{strlen} interceptor. \\

07 & \path{zend_mm_panic} &
\path{Zend/zend_alloc.c} &
3 &
\makecell[l]{\texttt{a3a529411e00}} &
\makecell[c]{7.0} &
SEGV during allocator panic. \\

08 & \path{zend_mm_panic} &
\path{Zend/zend_hash.c} &
1 &
\makecell[l]{\texttt{fb4e3085cbaa}} &
\makecell[c]{7.0} &
SEGV during heap free. \\

09 & \path{zend_call_function} &
\path{Zend/zend_execute_API.c} &
2 &
\makecell[l]{\texttt{3fe698b90493}} &
\makecell[c]{7.4} &
Stack overflow in call dispatch. \\

10 & \path{__interceptor_memcpy} &
\path{ext/mbstring/php_unicode.c} &
2 &
\makecell[l]{\texttt{2d48d734a201}} &
\makecell[c]{7.3} &
Stack-buffer-overflow in \texttt{memcpy}. \\

11 & \path{__interceptor_memcpy} &
\path{Zend/zend_smart_string.h} &
2 &
\makecell[l]{\texttt{e23a9d6a5c7d}\\\texttt{08755e6b8915}} &
\makecell[c]{7.3\\7.4} &
Stack overflow in \texttt{memcpy} interceptor. \\

12 & \path{__asan_region_is_poisoned} &
\path{Zend/zend_smart_string.h} &
1 &
\makecell[l]{\texttt{3fe698b90493}} &
\makecell[c]{7.4} &
Stack overflow in ASan poisoning check. \\

13 & \path{mbfl_convert_filter_get_vtbl} &
\path{ext/mbstring/libmbfl/mbfl/mbfl_convert.c} &
1 &
\makecell[l]{\texttt{8bb7417ca26c}} &
\makecell[c]{8.1} &
SEGV due to NULL vtable. \\

14 & \path{zval_addref_p} &
\path{Zend/zend_types.h} &
98 &
\makecell[l]{\texttt{7cf1450df14b}\\\texttt{a3a529411e00}} &
\makecell[c]{7.0} &
ABRT during refcount update. \\

15 & \path{_zval_get_long_func} &
\path{Zend/zend_operators.c} &
1 &
\makecell[l]{\texttt{caf007cdd604}} &
\makecell[c]{7.1} &
ABRT in \texttt{raise}. \\

16 & \path{convert_to_long_base} &
\path{Zend/zend_operators.c} &
1 &
\makecell[l]{\texttt{7cf1450df14b}} &
\makecell[c]{7.0} &
ABRT in \texttt{raise}. \\

17 & \path{mbfl_strlen} &
\path{ext/mbstring/libmbfl/mbfl/mbfilter.c} &
6 &
\makecell[l]{\texttt{2aceb0b00e21}\\\texttt{df2414a2230d}} &
\makecell[c]{8.2} &
ABRT in \texttt{raise}. \\

18 & \path{zif_range} &
\path{ext/standard/array.c} &
2 &
\makecell[l]{\texttt{bdc96afbdd25}\\\texttt{fb4e3085cbaa}} &
\makecell[c]{7.0} &
SEGV in range generation. \\

19 & \path{_zend_hash_del_el} &
\path{Zend/zend_hash.c} &
4 &
\makecell[l]{\texttt{bdc96afbdd25}} &
\makecell[c]{7.0} &
SEGV in hash deletion. \\

20 & \path{zend_assign_to_string_offset} &
\path{Zend/zend_execute.c} &
2 &
\makecell[l]{\texttt{7553c696c306}\\\texttt{e7123ef569d2}} &
\makecell[c]{8.0\\8.1} &
SEGV in string offset assignment. \\

21 & \path{zend_compile_unset} &
\path{Zend/zend_compile.c} &
1 &
\makecell[l]{\texttt{a3a529411e00}} &
\makecell[c]{7.0} &
ABRT during \texttt{unset} compilation. \\

22 & \path{zend_mm_free_heap} &
\path{Zend/zend_alloc.c} &
1 &
\makecell[l]{\texttt{8bb7417ca26c}} &
\makecell[c]{8.1} &
SEGV in \path{zend_mm_free_heap}. \\

23 & \path{_zendi_try_convert_scalar_to_number} &
\path{Zend/zend_operators.c} &
7 &
\makecell[l]{\texttt{57a4a2c5a8dd}\\\texttt{7553c696c306}\\\texttt{d54e197baee6}} &
\makecell[c]{8.0\\8.1} &
ABRT in \texttt{raise}. \\

24 & \path{zend_mm_alloc_small} &
\path{Zend/zend_alloc.c} &
1 &
\makecell[l]{\texttt{08755e6b8915}} &
\makecell[c]{7.3} &
Stack overflow in allocator. \\

\bottomrule
\end{longtable}
\end{eyeqapptable}
\par\medskip

\subsection{Unique crashes in RQ3}

\Cref{tab:rq3-fuzzing} presents all unique crashes found by \eyeqllm{} in the experiments for RQ3.

\begin{eyeqapptable}{4pt}
\begin{longtable}{
>{\centering\arraybackslash}p{0.65cm}  
>{\RaggedRight\arraybackslash}p{3.2cm} 
>{\RaggedRight\arraybackslash}p{3.6cm} 
>{\centering\arraybackslash}p{0.9cm}   
>{\centering\arraybackslash}p{1.35cm}  
>{\RaggedRight\arraybackslash}p{\dimexpr\textwidth-9.70cm-12\tabcolsep\relax} 
}
\caption{RQ3 --- 30 unique crashes found by \eyeqllm{} on the Sep-Dec 2025 code review dataset of PHP.}%
\label{tab:rq3-fuzzing}\\
\toprule
\textbf{ID} & \textbf{Function} & \textbf{Source} & \textbf{Occ.} & \textbf{PHP Ver.} & \textbf{ASAN Summary} \\
\midrule
\endfirsthead
\toprule
\textbf{ID} & \textbf{Function} & \textbf{Source} & \textbf{Occ.} & \textbf{PHP Ver.} & \textbf{ASAN Summary} \\
\midrule
\endhead
\bottomrule
\endfoot
\bottomrule
\endlastfoot

01 & \path{zend_objects_destroy_object} &
\path{Zend/zend_objects.c} &
318 &
\makecell[c]{8.4.14-dev\\8.5.0-dev\\8.6.0-dev} &
SEGV in \codeid{zend_objects_destroy_object}. \\

02 & \path{php_pcre_create_match_data} &
\path{ext/pcre/php_pcre.c} &
165 &
\makecell[c]{8.1.34-dev\\8.4.0-dev\\8.5.0-dev} &
ABRT in \texttt{raise}. \\

03 & \path{zend_reference_destroy} &
\path{Zend/zend_variables.c} &
15 &
\makecell[c]{8.5.0-dev} &
ABRT in \texttt{raise}. \\

04 & \path{ZEND_ASSIGN_OBJ_SPEC_VAR_TMPVAR_OP_DATA_TMP_HANDLER} &
\path{Zend/zend_vm_execute.h} &
15 &
\makecell[c]{8.5.0-dev} &
SEGV in VM handler. \\

05 & \path{zif_mb_str_pad} &
\path{ext/mbstring/mbstring.c} &
10 &
\makecell[c]{8.3.26-dev\\8.3.27-dev\\8.3.28-dev\\8.3.29-dev\\8.4.0-dev} &
FPE in \codeid{zif_mb_str_pad}. \\

06 & \path{zendi_try_get_long} &
\path{Zend/zend_operators.c} &
10 &
\makecell[c]{8.3.26-dev\\8.3.27-dev\\8.3.28-dev\\8.3.29-dev\\8.4.0-dev} &
ABRT in \texttt{raise}. \\

07 & \path{ZEND_POST_INC_OBJ_SPEC_VAR_CV_HANDLER} &
\path{Zend/zend_vm_execute.h} &
8 &
\makecell[c]{8.5.0-dev} &
SEGV in VM handler. \\

08 & \path{php_mb_check_encoding} &
\path{ext/mbstring/mbstring.c} &
6 &
\makecell[c]{8.3.27-dev\\8.3.28-dev\\8.3.29-dev} &
SEGV in \codeid{php_mb_check_encoding}. \\

09 & \path{strlen} &
\path{main/spprintf.c} &
8 &
\makecell[c]{8.3.28-dev\\8.5.0-dev\\8.6.0-dev} &
SEGV in \texttt{strlen}. \\

10 & \path{zend_mm_panic} &
\path{Zend/zend_alloc.c} &
3 &
\makecell[c]{8.4.0-dev} &
SEGV in allocator panic (\texttt{kill}). \\

11 & \path{_zend_is_inconsistent} &
\path{Zend/zend_hash.c} &
3 &
\makecell[c]{8.3.27-dev\\8.3.29-dev\\8.4.14-dev} &
ABRT in \texttt{raise}. \\

12 & \path{strlen} &
\path{ext/standard/mail.c} &
2 &
\makecell[c]{8.6.0-dev} &
SEGV in \texttt{strlen}. \\

13 & \path{_php_mb_regex_ereg_replace_exec} &
\path{ext/mbstring/php_mbregex.c} &
2 &
\makecell[c]{8.3.27-dev\\8.5.0-dev} &
ABRT in \texttt{raise}. \\

14 & \path{ZEND_PRE_INC_OBJ_SPEC_VAR_CV_HANDLER} &
\path{Zend/zend_vm_execute.h} &
2 &
\makecell[c]{8.5.0-dev} &
SEGV in VM handler. \\

15 & \path{onig_number_of_names} &
\path{ext/mbstring/php_mbregex.c} &
2 &
\makecell[c]{8.3.27-dev} &
SEGV in \codeid{onig_number_of_names}. \\

16 & \path{zend_mm_set_next_free_slot} &
\path{Zend/zend_alloc.c} &
2 &
\makecell[c]{8.5.0-dev} &
Stack overflow in \codeid{zend_mm_set_next_free_slot}. \\

17 & \path{zend_array_destroy} &
\path{Zend/zend_hash.c} &
3 &
\makecell[c]{8.4.0-dev\\8.5.0-dev} &
Stack overflow (also ABRT in \texttt{raise} in some cases). \\

18 & \path{zend_lazy_object_get_info} &
\path{Zend/zend_lazy_objects.c} &
1 &
\makecell[c]{8.5.0-dev} &
ABRT in \texttt{raise}. \\

19 & \path{zend_fetch_property_address} &
\path{Zend/zend_execute.c} &
1 &
\makecell[c]{8.5.0-dev} &
SEGV in \codeid{zend_fetch_property_address}. \\

20 & \path{zend_ast_evaluate_inner} &
\path{Zend/zend_ast.c} &
1 &
\makecell[c]{8.4.0-dev} &
Stack overflow in \codeid{zend_ast_evaluate_inner}. \\

21 & \path{strlen} &
\path{Zend/zend_execute.c} &
1 &
\makecell[c]{8.5.0-dev} &
SEGV in \texttt{strlen}. \\

22 & \path{zend_mm_free_small} &
\path{Zend/zend_alloc.c} &
2 &
\makecell[c]{8.5.0-dev} &
Stack overflow in \codeid{zend_mm_free_small}. \\

23 & \path{zend_mm_set_next_free_slot} &
\path{Zend/zend_hash.c} &
3 &
\makecell[c]{8.6.0-dev} &
Stack overflow in \codeid{zend_mm_set_next_free_slot}. \\

24 & \path{gc_check_possible_root} &
\path{Zend/zend_hash.c} &
1 &
\makecell[c]{8.6.0-dev} &
Stack overflow in \codeid{gc_check_possible_root}. \\

25 & \path{zend_string_release_ex} &
\path{Zend/zend_string.h} &
1 &
\makecell[c]{8.6.0-dev} &
SEGV in \codeid{zend_string_release_ex}. \\

26 & \path{zend_mm_free_small} &
\path{Zend/zend_hash.c} &
1 &
\makecell[c]{8.6.0-dev} &
Stack overflow in \codeid{zend_mm_free_small}. \\

27 & \path{zend_mm_free_heap} &
\path{Zend/zend_alloc.c} &
1 &
\makecell[c]{8.5.0-dev} &
Stack overflow in \codeid{zend_mm_free_heap}. \\

28 & \path{zend_mm_free_heap} &
\path{Zend/zend_hash.c} &
1 &
\makecell[c]{8.6.0-dev} &
Stack overflow in \codeid{zend_mm_free_heap}. \\

29 & \path{zend_array_destroy} &
\path{Zend/zend_execute.c} &
1 &
\makecell[c]{8.6.0-dev} &
Stack overflow in \codeid{zend_array_destroy}. \\

30 & \path{_emalloc} &
\path{Zend/zend_alloc.c} &
1 &
\makecell[c]{8.3.27-dev} &
Stack overflow in \codeid{_emalloc}. \\

\bottomrule
\end{longtable}
\end{eyeqapptable}
\par\medskip

\subsection{Previously unknown bugs}

\Cref{tab:bug-finding-appx-all} presents all previously unknown bugs discovered by \eyeq{}.

\begin{eyeqapptable}{2.2pt}
\begin{longtable}{
>{\centering\arraybackslash}p{0.6cm}  
>{\raggedright\arraybackslash}p{2.05cm} 
>{\centering\arraybackslash}p{0.95cm}  
>{\centering\arraybackslash}p{0.7cm}  
>{\centering\arraybackslash}p{0.95cm}  
>{\RaggedRight\arraybackslash}p{\dimexpr\textwidth-5.25cm-12\tabcolsep\relax} 
}
\caption{All previously unknown bugs discovered by \eyeq{} in our experiments. ``Cf'' indicates the bug has been confirmed by PHP core developers, ``Fx'' denotes a bug that has been fixed upstream, and ``Pd'' means a bug is pending for further analysis. A missing Issue ID or Description (``-'') denotes a security bug whose details are intentionally withheld to comply with responsible disclosure.}%
\label{tab:bug-finding-appx-all}\\
\toprule
\textbf{ID} & \textbf{Crash Location} & \textbf{Issue ID} & \textbf{Status} & \textbf{Fixes} & \textbf{Description} \\
\midrule
\endfirsthead
\toprule
\textbf{ID} & \textbf{Crash Location} & \textbf{Issue ID} & \textbf{Status} & \textbf{Fixes} & \textbf{Description} \\
\midrule
\endhead
\bottomrule
\endfoot
\bottomrule
\endlastfoot
01 & \path{zend_hash.c} & \issuelink{19613} & Fx & \textcolor{ForestGreen}{+31} \textcolor{red}{-2} & SEGV caused by incorrect iterator reset handling during Copy-on-Write array modification. \\
02 & \path{zend_hash.c} & \issuelink{19839} & Fx & \textcolor{ForestGreen}{+22} \textcolor{red}{-1} & Incorrect \codeid{HASH_FLAG_HAS_EMPTY_IND} flag on userland array. \\
03 & \path{zend_hash.c} & \issuelink{19840} & Fx & \textcolor{ForestGreen}{+64} \textcolor{red}{-10} & Hash table corruption during shutdown when bailing out of user stream handling. \\
04 & \path{zend_hash.c} & \issuelink{20286} & Fx & \textcolor{ForestGreen}{+33} \textcolor{red}{-13} & Hash table use-after-destroy in function table during stream wrapper cleanup. \\
05 & \path{zend_hash.c} & \issuelink{19605} & Cf & - & Failed reference count assertion in \codeid{zend_hash_index_add_or_update_i()} during table update. \\

06 & \path{zend_hash.c} & \issuelink{20503} & Fx & \textcolor{ForestGreen}{+49} \textcolor{red}{-0} & Assertion failure in hash-table property-hash construction when using \texttt{DateInterval} in complex nested assignments/casts, leading to crash in \codeid{_zend_hash_str_add_or_update_i()}. \\

07 & \path{zend_hash.c} & \issuelink{20477} & Cf & - &
Heap use-after-free in \codeid{zend_array_count} (Zend Hash), calling \codeid{zend_array_count} on a freed HashTable triggers invalid memory access. \\

08 & \path{zend_hash.c} & \issuelink{20656} & Cf & - &
HashTable use-after-destroy in function table triggered by stream-wrapper magic via nested redeclaration in \codeid{__call()}; crash arises inside \codeid{zend_hash} during cleanup. \\

09 & \path{zend_hash.c} & \issuelink{20855} & Cf & - &
Failure in \codeid{zend_hash_del} with HashTable refcount inconsistency. \\

10 & \path{spl_dllist.c} & \issuelink{20856} & Fx & \textcolor{ForestGreen}{+35} \textcolor{red}{-2} &
SPL: heap-use-after-free in SplDoublyLinkedList iterator when modifying during iteration. \\

11 & \path{basic_functions.c} & \issuelink{20906} & Fx & \textcolor{ForestGreen}{+79} \textcolor{red}{-3} &
Assertion failure in \codeid{zif_highlight_string} when messing up output buffers. \\

12 & \path{mail.c} & \issuelink{20858} & Fx & \textcolor{ForestGreen}{+47} \textcolor{red}{-1} &
Null pointer dereference in \codeid{php_mail_detect_multiple_crlf} via \codeid{error_log}. \\

13 & \path{url.c} & \issuelink{20088} & Fx & \textcolor{ForestGreen}{+1155} \textcolor{red}{-31} & Heap UAF in WHATWG URI parser during malformed URL processing. \\

14 & \path{url.c} & \issuelink{20502} & Fx & \textcolor{ForestGreen}{+19} \textcolor{red}{-10} &
SEGV in WHATWG URL parser (Lexbor integration) when parsing malformed URLs, memory corruption in URL handling. \\

15 & \path{output.c} & \issuelink{20352} & Fx & \textcolor{ForestGreen}{+34} \textcolor{red}{-4} & Heap UAF in \codeid{php_output_handler_free} via re-entrant \codeid{ob_start()} during error deactivation. \\

16 & \path{zend_compile.c} & \issuelink{20113} & Fx & \textcolor{ForestGreen}{+17} \textcolor{red}{-1} & Missing \texttt{new} expression handling in constant expressions. \\

17 & \path{zend_gc.c} & \issuelink{19983} & Cf & - & GC failure due to unaligned reference pointer in \codeid{gc_collect_roots()}. \\

18 & \path{zend_gc.h} & \issuelink{20861} & Cf & - & GC failure in \codeid{gc_check_possible_root_no_ref} with lazy proxy. \\

19 & \path{zend_types.h} & \issuelink{19999} & Cf & - & GC reference counting assertion failure via invalid object assignment. \\

20 & \path{zend_types.h} & \issuelink{20519} & Cf & - & Use-after-destroy in type-system structures triggered by nested magic \codeid{__call()} inside stream wrappers, causing reuse of freed entries and potential memory corruption. \\

21 & \path{zendlex.c} & \issuelink{20564} & Fx & \textcolor{ForestGreen}{+32} \textcolor{red}{-63} & Failure in \texttt{zendlex()} during re-entrant \codeid{__call()} execution triggered by \texttt{eval()} + autoloader registration, can lead to interpreter abort or memory corruption. \\

22 & \path{zend_fibers.c} & \issuelink{20483} & Fx & \textcolor{ForestGreen}{+26} \textcolor{red}{-1} &
Stack-overflow in \codeid{zend_fiber_execute()} when using a too-small \codeid{fiber.stack_size} (invalid ``\codeid{fiber.stack_size}'' setting), leading to crash/stack overflow. \\

23 & \path{zend_fibers.c} & \issuelink{20615} & Cf & - &
Failure in \codeid{check_node_running_in_fiber()} when resuming a generator inside a Fiber destructor, leading to crash during fiber cleanup. \\

24 & \path{php_reflection.c} & \issuelink{20174} & Fx & \textcolor{ForestGreen}{+93} \textcolor{red}{-42} & Failure in \texttt{ReflectionProperty::skipLazyInitialization} after failed LazyProxy initialization. \\

25 & \path{zend_object_handlers.c} & \issuelink{20177} & Fx & \textcolor{ForestGreen}{+32} \textcolor{red}{-0} & Incorrect handling of overridden private properties in object variables. \\

26 & \path{zend_object_handlers.c} & \issuelink{20479} & Fx & \textcolor{ForestGreen}{+42} \textcolor{red}{-1} & Heap buffer overflow in property-hook logic: unbounded append in \codeid{zho_build_properties_ex()} when a re-entrant getter is triggered during \codeid{var_export()}, leading to memory corruption. \\

27 & \path{zend.c, ...} & - & Cf & - & - \\

28 & \path{zend_objects.c} & \issuelink{20183} & Fx & \textcolor{ForestGreen}{+73} \textcolor{red}{-3} & Heap UAF via stale \codeid{EG(opline_before_exception)} pointer during \texttt{eval}. \\

29 & \path{zend_objects.c} & \issuelink{20324} & Fx & \textcolor{ForestGreen}{+45} \textcolor{red}{-13} & NULL \codeid{EG(current_execute_data)} during destructor execution. \\

30 & \path{zend_objects.c} & - & Pd & - & - \\

31 & \path{zend_objects.c} & \issuelink{20612} & Cf & - &
HashTable use-after-destroy during lazy-proxy \codeid{__get()} with dynamic properties. \\

32 & \path{zend_objects.c} & \issuelink{20657} & Fx & \textcolor{ForestGreen}{+26} \textcolor{red}{-1} &
Assertion failure in \codeid{zend_lazy_object_get_info} triggered by calls to \texttt{setRawValueWithoutLazyInitialization()} and \texttt{newLazyGhost()} on lazy-loaded objects, leading to crash when inspecting object info. \\

33 & \path{zend_lazy_objects.c} & \issuelink{20905} & Fx & \textcolor{ForestGreen}{+24} \textcolor{red}{-2} &
Assertion failure in \codeid{zend_lazy_object_del_info} when cloning lazy proxy. \\

34 & \path{zend_execute.c} & \issuelink{20270} & Fx & \textcolor{ForestGreen}{+55} \textcolor{red}{-1} & NULL pointer dereference when handling named arguments in property hook setter. \\

35 & \path{zend_execute.h} & \issuelink{20614} & Fx &
\textcolor{ForestGreen}{+29} \textcolor{red}{-2} &
Incorrect reference-handling during unserialization of \texttt{SplFixedArray} causes an assertion failure in the Zend VM execution layer; resolved by fixing reference propagation in \codeid{zend_execute.h}. \\

36 & \path{zend_vm_execute.h} & \issuelink{20073} & Fx & \textcolor{ForestGreen}{+16} \textcolor{red}{-1} & Failure in WeakMap offset operations on reference. \\

37 & \path{zend_vm_execute.h} & \issuelink{20085} & Fx & \textcolor{ForestGreen}{+29} \textcolor{red}{-2} & Missing exception handling in \codeid{FETCH_RW} handler. \\

38 & \path{zend_vm_execute.h} & \issuelink{20171} & Cf & - & Failure during LazyProxy initialization with inherited by-reference \codeid{__get()} method. \\

39 & \path{zend_vm_execute.h} & \issuelink{20482} & Cf & - & Use-after-free in VM: object slot reused after exception thrown in object instantiation, leading to potential memory corruption on subsequent variable access. \\

40 & \path{zend_ini_parser.y} & \issuelink{20695} & Fx & \textcolor{ForestGreen}{+15} \textcolor{red}{-1} &Uninitialized memory in PHP's INI string parser (\codeid{parse_ini_string}). \\

41 & \path{php_dom.c} & \issuelink{20722} & Fx & \textcolor{ForestGreen}{+20} \textcolor{red}{-5} &Null pointer dereference in DOM namespace node cloning via clone on malformed objects. \\

\midrule

42 & \path{php_mbregex.c} & \issuelink{21036} & Pd & -  &Null pointer dereference in \texttt{mb\_ereg\_search\_getregs()} after \texttt{mb\_eregi()} invalidates regex cache. \\

43 & \path{mbstring.c} & \issuelink{20833} & Fx & \textcolor{ForestGreen}{+25} \textcolor{red}{-0}  & \texttt{mb\_str\_pad()} divide by zero if padding string is invalid in the encoding \\

44 & \path{zend_hash.c} & \issuelink{20836} & Fx & \textcolor{ForestGreen}{+127} \textcolor{red}{-13}  & Stack overflow in mbstring variable conversion with recursive array references. \\

45 & \path{zend_ast.c} & \issuelink{21072} & Fx & \textcolor{ForestGreen}{+22} \textcolor{red}{-0}  & \texttt{zend\_ast} assertion failure in class property initializer. \\

46 & \path{php_mbregex.c} & - & Cf & - & - \\
\bottomrule
\end{longtable}
\end{eyeqapptable}

\end{document}